\documentclass[longbibliography,showpacs,aps,prd,twocolumn,floatfix]{revtex4-1}

\usepackage{array}
\usepackage{color}
\usepackage{graphicx}
\usepackage{dcolumn}
\usepackage{bm}
\usepackage{gensymb}
\usepackage{mathcomp}
\usepackage{textcomp}
\usepackage{setspace}
\usepackage{amsmath}
\usepackage{amsthm}
\usepackage{amssymb}
\usepackage{mathtools}
\usepackage{float}

\DeclareMathOperator*{\argmin}{arg\,min}
\newcommand{\scR}{\mathcal{R}}
\newcommand{\scL}{\mathcal{L}}
\newcommand{\scRd}{\mathcal{R}_{\text{direct}}}
\newcommand{\scLd}{\mathcal{L}_{\text{direct}}}

\newcommand{\pFq}[2]{{}_{#1}F_{#2}}

\newcommand{\dimnabla}{\nabla}
\newcommand{\dimrho}{\widetilde{\rho}}
\newcommand{\dimphi}{\phi}
\newcommand{\dimpsi}{\psi}
\newcommand{\ndrho}{\rho}
\newcommand{\ndphi}{\Phi}
\newcommand{\ndpsi}{\Psi}

\newcommand{\dimphiASE}{\dimphi_{\textrm{SE}}}
\newcommand{\dimphiSpE}{\dimphi_{\textrm{S+E}}}
\newcommand{\dimphiS}{\dimphi_{\textrm{S}}}
\newcommand{\dimphiSEM}{\dimphi_{\textrm{SEM}}}

\newcommand{\dimphiEM}{\dimphi_{\textrm{EM}}}
\newcommand{\dimphiSpEpM}{\dimphi_{\textrm{S+E+M}}}
\newcommand{\dimpsiG}{\dimpsi_{\textrm{G}}}

\newcommand{\rref}{r_{\textrm{ref}}}
\newcommand{\Rref}{R_{\textrm{ref}}}
\newcommand{\dimphiref}{\dimphi_{\textrm{ref}}}
\newcommand{\dimrhoref}{\rho_{\textrm{ref}}}
\newcommand{\dimMref}{M_{\textrm{ref}}}

\begin{document}

\title{Robust numerical computation of the 3D scalar potential field \\ of the cubic Galileon gravity model at solar system scales}
\author{Nicholas C. White}
\email[Corresponding author: ]{nwhite@caltech.edu}
\author{Sandra M. Troian}
\affiliation{\\
T. J. Watson Sr. Laboratories of Applied Physics - MC 128-95 \\
California Institute of Technology\\Pasadena, CA 91125, USA}
\author{Jeffrey B. Jewell}
\author{Curt J. Cutler}
\author{Sheng-wey Chiow}
\author{Nan Yu}
\affiliation{Jet Propulsion Laboratory\\California Institute of Technology\\4800 Oak Grove Drive,
Pasadena, CA 91109}
\date{\today}

\begin{abstract}
 Direct detection of dark energy or modified gravity may finally be within reach due to ultrasensitive instrumentation such as atom interferometry capable of detecting incredibly small scale accelerations. Forecasts, constraints and measurement bounds can now too perhaps be estimated from accurate numerical simulations of the fifth force and its Laplacian field at solar system scales. The cubic Galileon gravity scalar field model (CGG), which derives from the DGP braneworld model, describes  modified gravity incorporating a Vainshtein screening mechanism. The nonlinear derivative interactions in the CGG equation suppress the field near regions of high density, thereby restoring general relativity (GR) while far from such regions, field enhancement is comparable to GR and the equation is dominated by a linear term. This feature of the governing PDE poses some numerical challenges for computation of the scalar potential, force and Laplacian fields even under stationary conditions. Here we present a numerical method based on finite differences for solution of the static CGG scalar field for a 2D axisymmetric Sun-Earth system and a 3D Cartesian Sun-Earth-Moon system. The method relies on gradient descent of an integrated residual based on the normal attractive branch of the CGG equation. The algorithm is shown to be stable, accurate and rapidly convergent toward the global minimum state. We hope this numerical study, which can easily be extended to include smaller bodies such as detection satellites, will prove useful to future measurement of modified gravity force fields at solar system scales.
\end{abstract}
\date{\today}

\maketitle

\section{Introduction}
\label{sec:introduction}
Researchers continue to debate whether dark energy or modified gravity is responsible for the apparent accelerating expansion of the universe. It has been proposed that the source of this debate involves our fundamental lack of understanding of gravitational physics at large scales \cite{R12,KS07,Sch09a,SL10}. To accord with the observed lack of a strong ``fifth force'' on bodies at sub-solar-system scales, these models typically include a nonlinear screening mechanism whereby the field value in the vicinity of dense compact bodies is suppressed and GR restored, while at large distances from dense bodies the field is comparable to GR and the equation dominated by a linear term. Current proposals include the Chameleon mechanism \cite{KW04}, the Symmetron mechanism \cite{OP08,HK10}, and the Vainshtein mechanism \cite{V72}, the last of which arises in a variety of Galileon models \cite{NT09,R12}. In this work, we focus exclusively on the cubic Galileon gravity (CGG) form of the Vainshtein mechanism, which arises in models including the Dvali-Gabadadze-Porrati (DGP) braneworld model \cite{DGP00}.

In scalar field models that incorporate Vainshtein screening, the standard quadratic kinetic term in the Lagrangian is augmented by a higher-order nonlinearity and the system is forced by the trace of the stress-energy tensor. The resulting governing equation includes a linear d'Alembertian term, which dominates at long distances (based on a length scale derived from the forcing of the given system), and a nonlinear term, which suppresses the field at small distances \cite{R12}. A particularly simple manifestation of the Vainshtein mechanism, and the one investigated in this study, is one in which the Lagrangian contains a cubic interaction term, hence leading to an equation that is quadratic in the field and higher derivatives.

One notable example giving rise to such a cubic Lagrangian is the DGP model, in which the universe is regarded as a 4D brane embedded in a 5D Minkowski bulk space. The corresponding action contains two Einstein-Hilbert terms, one for the bulk and one for the brane, each with its own Planck mass setting the strength of the gravity, which may be denoted $M_4$ and $M_5$, respectively. These two mass scales induce a crossover length scale $r_c = (\hbar/c) M_4^2/\left(2 M_5^3\right)$, where $\hbar$ denotes Planck's constant and $c$ is the speed of light, which characterizes the distance over which metric fluctuations propagating on the brane dissipate into the bulk, above which 5D gravity is dominant and below which 4D gravity is dominant \cite{DGP00}. Metric perturbations can then be linearized as scalar fields acting on the brane in the decoupling limit. These perturbations to the metric describe leakage of gravitons from the 4D brane universe \cite{LS03}. The resulting equation of motion is dominated by linear terms at scales above the Vainshtein radius $r_V$, and nonlinear terms below $r_V$, where $r_V \sim r_c^{1/3}$ and is also dependent on details of the local density. To make contact with current cosmological models, $r_c$ is typically expressed in terms of the current Hubble rate $H_0$ and matter density parameter $\Omega_{m}^{0}$ such that $r_c = c H_0^{-1}\left( 1 - \Omega_m^0 \right)^{-1}$ \cite{D01,CS09}. The crossover length evaluates to approximately $1.8 \times 10^{23} \, \text{km}$ or $6 \times 10^3 \, \text{Mpc}$ using the constants in Ref. [\onlinecite{CO17}].

In the DGP model there is an unstable self-accelerating branch and a stable non-accelerating attractive branch \cite{D01,CP06}; the present work considers only the latter. Scalar field models such as these are typically constrained by comparing astronomical measurements to their predictions at cosmological or galactic scales (e.g., Mpc scales) at which the field is still relatively strong. Such predictions are often based on density perturbation analyses and n-body simulations. So far, analytic approaches have mostly yielded field equations for large-scale density perturbations \cite{D02,LS04,KM06,KS07,Sco09,SL10}. Numerical simulations of large-scale structure evolution in the quasi-static approximation have also been conducted using spectral \cite{CS09} and position-space \cite{Sch09a,LK13,BP13} methods which sequentially solve for the scalar field and update the local mass density distribution, represented as discrete particles. These large-scale simulations have predicted perturbative density growth rates, power spectra of mass distribution, and parameter values for dark matter halos. Besides large-scale structure formation, the dynamics and radiation of binary pulsars under a DGP-like scalar field have also been studied and shown to influence orbital periods due to Vainshtein screening \cite{RT13,RW13,DT18}. Analysis of the dependence of the Vainshtein radius on the radii of bodies has also demonstrated that the relative strength of the cubic Galileon fifth force to gravity is greater around infinite cylindrical bodies than around spherical bodies \cite{BD15}, indicating potential advantages in obtaining measurements in regions lacking spherical symmetry in order to better discriminate signals from the fifth force. 

On the experimental side, advances in instrumentation, such as atom interferometry, have introduced unprecedented sensitivity in force measurements \cite{CY15,WM16,CY16}, so much so that there now exists the possibility of direct detection of a ``fifth force'' due to modified gravity; indeed, detection schemes for Chameleon \cite{BH15, HK15, CY18} and Symmetron \cite{CY19} models have already been proposed. To determine bounds on measurement precision and to support mission concept studies for direct detection of the cubic Galileon scalar field, accurate solution of the scalar potential field at solar system scales is therefore required.

Numerical investigation of the 3D Galileon potential field at solar system scales described by the CGG model carries some inherent challenges. In contrast to the behavior at large-scales, the linear term at small distances is essentially negligible, such that the field equation becomes strongly nonlinear in the second derivative terms. Methods based on the finite element technique therefore become difficult to apply. Unlike the large-scale n-body regime, the solar system regime contains mass sources with compact support, such as the Sun and planets, which introduces difficulties for spectral methods, and further suggests modeling mass density as a field rather than as discrete particles. And since the radii of  bodies tend to be orders of magnitude smaller than their separation distances, the multiple scales inherent in this system must be managed effectively to prevent numerical artifacts. Furthermore, since the CGG equation is quadratic, it harbors both attractive and repulsive solutions; care must be taken in isolating solutions iterating toward two separate global minima. Despite these challenges, some numerical studies have successfully elucidated aspects at small scales within the Vainshtein radii of the relevant bodies. For example, the anomalous precession of bodies such as Mercury beyond the correction to GR has been computed \cite{LS03,I12} as has solution of the Green's functions for corrections to a massive, spherically-symmetric body, with  perturbative corrections computed to several orders \cite{AT13,CT13}.

While prior work had concentrated on large-scale cosmological simulations, Hiramatsu \textit{et al.} \cite{HS13} realized the importance of studying the very different small-scale regime as well and carried out the first significant numerical study of the static scalar potential field equation at small scales. Their study considered an idealized system containing two spherical bodies with mass ratio comparable to Earth and the Moon but positioned within very close range, using a finite difference technique coupled with a successive over-relaxation method. The system considered was within the nonlinear regime subject to strong screening since the two bodies were both well within each other's Vainshtein radius. In this case, the authors were able to take advantage of the fact that despite the presence of a strong nonlinear term, the solution at distances close to a massive body must be dominated by that body. (We note the use of the term  ``screening'' in Ref. \cite{HS13} to describe this effect should not be confused with the Vainshtein screening mechanism). Recent studies have also examined masses contained within spherical shells or voids which become subject to a force, in contrast to masses subject purely to Newtonian gravity \cite{BH13}. Numerical simulations of disks containing holes have revealed how cavities reduce the screening force \cite{OK19}. In these examples, the numerical iteration scheme converges well so long as the initial trial solution is sufficiently close to the true solution.

The goal of this current work therefore is to provide an accurate and rapidly convergent numerical scheme for solution of the static scalar potential field of the DGP braneworld model with cubic Galileon interaction at solar system scales for systems containing multiple dense compact mass sources. We present a numerical method based on finite differences for solution of the static CGG scalar field for a 2D axisymmetric Sun-Earth system and a 3D Cartesian Sun-Earth-Moon system. The method relies on gradient descent of an integrated residual based on the normal attractive branch of the CGG equation. The algorithm is shown to be stable, accurate and rapidly convergent toward the global minimum state.

This paper is organized as follows: Section \ref{sec:analytic} describes the model system and non-dimensionalization of the governing equation to identify dominant and subdominant terms; Section \ref{sec:results} outlines the iteration scheme with application to solution of the Galileon potential fields for the axisymmetric Sun-Earth system and 3D Sun-Earth-Moon system along with discussion of results. Following the conclusion, we provide in the Appendices detailed explanations of the numerical method along with validation tests. Included there are download links to the data and software for the interested reader. 

\section{Analytic model and rescalings}
\label{sec:analytic}
In what follows, we follow the derivations in Refs. \cite{SL10} and \cite{NR04} and review features of the CGG model equation. In particular, we discuss the known analytic solution for a spherically-symmetric single body at length scales above and far below the Vainshtein limit. The CGG equation are then non-dimensionalized to highlight relative strengths of the linear and nonlinear terms at solar system scales and to facilitate numerical investigation.

In the limit of large Planck masses, the helicity-2 and helicity-0 modes in the DGP model decouple so that the helicity-0 mode can be considered in isolation as a scalar field with effective action described by the cubic Lagrangian 
\begin{align}
  \label{eqn:MinkowskiLagrangian}
  \mathfrak{L} &= 3 \dimphi \Box \dimphi - \frac{r_c^2}{c^2} \left(\partial_\alpha \dimphi\right) \left(\partial^\alpha \dimphi\right) \Box\dimphi + 16 \pi G c^{-2} T^{\alpha}_{~\alpha} \dimphi
  ,
\end{align}
where $\Box$ is the d'Alembertian operator, $\partial_\alpha$ is the 4D covariant derivative, $\dimphi(\vec{r})$ is the scalar field, $G$ is the gravitational constant, and $T^{\alpha}_{~\alpha}$ is the trace of the stress-energy tensor \cite{NR04,R12}.
Eq. \ref{eqn:MinkowskiLagrangian} was derived for a Minkowski (flat) 4D space; in a Friedmann background, the Lagrangian can instead be written as
\begin{align}
  \label{eqn:FriedmannLagrangian}
  \mathfrak{L}_{\textrm{Fr}} &= 3 \frac{\beta}{a^2} \dimphi \Box \dimphi - \frac{r_c^2}{a^4 c^2} \left(\partial_\alpha \dimphi\right) \left(\partial^\alpha \dimphi\right) \Box\dimphi + 16 \pi G c^{-2} T^{\alpha}_{~\alpha} \dimphi
  ,
\end{align}
where $a$ is the cosmological scale factor and $\beta = 1 \pm 2 H_0 c^{-1} r_c [1 + (\partial_t H_0)/(3 H_0^2)]$, where $\partial_t$ denotes the time derivative. The positive branch of $\beta$ corresponds to the normal (stable) DGP branch \cite{KS07,SL10}. By definition, $a=1$ at the present day and, depending on the deceleration factor of the Universe, $\beta$ may be estimated by a value of 2--4. Following Ref. \cite{HS13}, we approximate $\beta \approx 1$ and proceed with the Lagrangian defined in Eq. (\ref{eqn:MinkowskiLagrangian}).

The resulting equation of motion is given by
\begin{align}
  3 \Box \dimphi + \frac{r_c^2}{c^2}\left[ \left( \Box \dimphi \right)^2 - \partial_{\alpha \mu} \dimphi \partial^{\alpha \mu} \dimphi \right] &= -8 \pi G c^{-2} T^{\alpha}_{~\alpha}
  ,
\end{align}
whose static potential field satisfies the nonlinear equation
\begin{align}
  \label{eqn:dimPotential}
  3 \dimnabla^2 \dimphi + \frac{r_c^2}{c^2}\left[ \left( \dimnabla^2 \dimphi \right)^2 - \sum_{i,j} \left(\dimnabla_{i}\dimnabla_{j} \dimphi\right)^2 \right] &= 8 \pi G \dimrho
  ,
\end{align}
where $\dimrho$ denotes the local mass density difference from the cosmological mean and $\dimnabla_i$ is the 3D gradient operator. We note henceforth that the Einstein summation convention no longer applies. The scalar field can be regarded as static since the configuration of solar system bodies changes very slowly in comparison with the speed of light. 

In the vicinity of dense bodies, the nonlinear term is dominant due to size of the coefficient $r_c$; at long scales above the Vainshtein radius, the linear term is dominant. The transition in solution behavior that results, is evident, for example, in the spherically-symmetric solution $\dimphi(r)$ for a single mass source, for which Eq. (\ref{eqn:dimPotential}) reduces to the form \cite{SL10}
\begin{align}
  \label{eqn:dimPotentialRadial}
  \frac{6}{r}\partial_r \dimphi + 3\partial_{rr} \dimphi + \frac{r_c^2}{c^2} \frac{2}{r^2}\left(\partial_r \dimphi\right)\left( \partial_r \dimphi + 2 r \partial_{rr} \dimphi \right) &= 8 \pi G \dimrho
  .
\end{align}
Schmidt \textit{et al.} \cite{SL10} showed that the solution $\dimphi(r)$ for a single spherical body of radius $\rref$, density $\dimrhoref$ and total mass $\dimMref = (4/3) \pi \dimrhoref \rref^3$ can be written in terms of the hypergeometric function $\pFq{2}{1}$, where
\begin{widetext}
\begin{align}
  \label{eqn:dimOnebodySoln}
  \dimphi(r) &= \frac{3 c^2}{8}\begin{dcases}
    \left( \frac{r}{r_c} \right)^2\left[ \sqrt{1 + \left( \frac{r_V}{\rref} \right)^{3}} - 1 \right] +
\left( \frac{\rref}{r_c} \right)^2
    \left(
      \pFq{2}{1}\left[-\frac{1}{2}, -\frac{2}{3}; \frac{1}{3}; -\left(\frac{r_V}{\rref}\right)^3\right]
    - \sqrt{1 + \left( \frac{r_V}{\rref} \right)^{3}} \right), &r \le \rref;
    \\
    \left( \frac{r}{r_c} \right)^2\left(
      \pFq{2}{1}\left[-\frac{1}{2}, -\frac{2}{3}; \frac{1}{3}; -\left(\frac{r_V}{r}\right)^3\right]
  - 1 \right), &r > \rref;
  \end{dcases}
\end{align}
\end{widetext}
and the Vainshtein radius is given by \cite{D02}
\begin{align}
  r_V = \frac{4}{3} \rref \left( \pi \frac{G}{c^2} \dimrhoref r_c^2 \right)^{1/3} = \left( \frac{16}{9} \frac{G}{c^2} \dimMref r_c^2 \right)^{1/3}
  .
\end{align}
The constant of integration incorporated into this form, which ensures $\lim \dimphi (r \to \infty) = 0$, does not affect the resulting force since the addition of a constant to the potential is a gauge freedom.

Given the complex nature of this solution, it is useful to examine instead the resulting force on a test body, which yields the simpler expression \cite{NR04,SL10}:
\begin{align}
  \partial_r \dimphi(r) &= \frac{3 c^2 r}{4 r_c^2} \begin{dcases}
    \sqrt{1 + \frac{64}{27} \pi \frac{G}{c^2} \dimrhoref r_c^2} - 1 ,&r \le \rref ,\\
    \sqrt{1 + \frac{64}{27} \pi \frac{G}{c^2} \dimrhoref r_c^2\left( \frac{\rref}{r} \right)^3} - 1 ,&r > \rref
    ,
  \end{dcases}
\end{align}
  or equivalently,
\begin{align}
  \partial_r \dimphi(r) &= \frac{3 c^2 r}{4 r_c^2} \begin{dcases}
    \sqrt{1 + \left( \frac{r_V}{\rref} \right)^3} - 1 ,&r \le \rref ,\\
    \sqrt{1 + \left( \frac{r_V}{r} \right)^3} - 1 ,&r > \rref
  .
  \end{dcases}
\end{align}
This solution corresponding to the positive square root identifies the attractive solution which correctly matches the $r \to \infty$ linear-dominated limit of Eq. (\ref{eqn:dimPotential}).

At short distances where $\rref < r \ll r_V$, wherein the nonlinear terms dominate, the first-order solution becomes \cite{NR04,SL10}:
\begin{align}
  \lim_{\rref < r \ll r_V}
  \partial_r \dimphi(r) = \frac{3 c^2 r_V}{4 r_c^2}\left[ \left( \frac{r}{r_V} \right)^{-1/2} + O\left( \frac{r}{r_V} \right) \right]
  .
\end{align}
At long distances $r \gg r_V$ in which the linear term is instead dominant, the solution to first order becomes harmonic and reduces to \cite{NR04,SL10}:
\begin{align}
  \lim_{r \gg r_V}
  \partial_r \dimphi(r) = \frac{3 c^2 r_V}{8 r_c^2}\left[ \left( \frac{r}{r_V} \right)^{-2} + O\left( \frac{r}{r_V} \right)^{-4} \right]
  .
\end{align}

\subsection{Non-dimensionalization of scalar potential equation}
\label{sec:analytic:nondim}
We introduce here scalings for non-dimensionalization of the governing scalar potential equation in order to clarify the relative importance of the linear term at small scales and to simplify the numerical method. The rescaling is based on a suitable length scale of interest, $d$, and a reference spherical mass of radius $\rref$ and density $\dimrhoref$. These choices yield a  characteristic scale for the potential field $\dimphiref$ and a dimensionless coefficient $k$ preceding the linear term, where
\begin{subequations}
\begin{align}
  \dimphiref &= \left( \frac{3}{2} \right)^{3/2}\frac{c^2 d^{1/2} r_V^{3/2}}{r_c^2}
  = \sqrt{\left( 8 \pi G \dimrhoref \right)\frac{d\,c^2\, \rref^{3}}{r_c^2}}
  ,
  \\
  k &= \sqrt{\frac{8}{3}} \left( \frac{d}{r_V} \right)^{3/2}
  = \sqrt{\frac{9}{8 \pi}\left(\frac{d^3\, c^2}{G \dimrhoref \rref^3 r_c^2} \right)}
  .
\end{align}
\end{subequations}
The resulting non-dimensional equation for the scalar field $\ndphi(\vec{R})$ becomes
\begin{align}
  \label{eqn:nd_vainshtein_pot}
  k \nabla^2 \ndphi +
  \left[ \left( \nabla^2 \ndphi \right)^2 - \sum_{i,j} \left(\nabla_{i}\nabla_{j} \ndphi\right)^2 \right]
  &= \ndrho
  ,
\end{align}
where the reduced density $\ndrho = \dimrho/(\dimrhoref \rref^3/d^3)$ and dimensionless reference body radius is $\Rref =\rref/d$. All other scalings and definitions can be found in Table \ref{table:simulationparams}.

The solution in Eq. (\ref{eqn:dimOnebodySoln}) can now be recast in dimensionless form for a spherically symmetric body of radius $R_b = r_b/d$ and density $\ndrho_b = \dimrho_b/(\dimrhoref \rref^3/d^3)$:
\begin{widetext}
\begin{align}
  \label{eqn:ndOnebodySoln}
  \ndphi(R) &=
  \frac{kR_b^2}{8}
  \begin{dcases}
    \left(\frac{R}{R_b}\right)^2\left[ \sqrt{1 + \frac{8 \ndrho_b}{3k^2}} - 1 \right] +
    \pFq{2}{1}\left[-\frac{1}{2}, -\frac{2}{3}; \frac{1}{3}; -\frac{8 \ndrho_b}{3k^2}\right]
    - \sqrt{1 + \frac{8\ndrho_b}{3k^2}},&R \le R_b;
    \\
    \left(\frac{R}{R_b}\right)^2 \left(
    \pFq{2}{1}\left[-\frac{1}{2}, -\frac{2}{3}; \frac{1}{3}; - \frac{8 \ndrho_b}{3 k^2} \left( \frac{R_b}{R} \right)^3 \right]
    - 1 \right),&R > R_b.
  \end{dcases}
\end{align}
\end{widetext}

In regions close to a dense mass such as a planet, the reference density $\dimrhoref$ will typically be tens of orders of magnitude greater than the cosmological average density \cite{R16}; hence even if the surrounding space has an underdensity, it will tend to be negligibly small. The empty space close to a dense mass may therefore be assumed to have a value $\ndrho \ge 0$.

The chosen scalings help distinguish between solutions characterized by $k$ large and $k$ small and in turn make evident whether the linear or nonlinear term in Eq. (\ref{eqn:nd_vainshtein_pot}) is dominant at a given distance. The scalings above derive from consideration of a single body with spherically symmetry, but may be applied to the case of multiple bodies by choosing an appropriate distance $d$ and either one body or a combination of the bodies for the reference mass and radius. In the case of one massive body which dominates the fields of all other bodies, such as the Sun in the solar system, the massive body is the natural reference choice. In the highly nonlinear regime characterized by negligibly small values of $k$, the governing equation for $\ndphi(\vec{R})$ retains only the nonlinear terms, thereby simplifying to the form
\begin{align}
  \label{eqn:vainshtein_nonlin_only}
  \left[ \left( \nabla^2 \ndphi \right)^2 - \sum_{i,j} \left(\nabla_{i}\nabla_{j} \ndphi\right)^2 \right]
  & = \ndrho
  .
\end{align}
The single-body spherically-symmetric solution in this nonlinear regime is then given by
\begin{align}
\label{eqn:spherk0}
  \ndphi(R) &=
  \sqrt{\frac{M_b R_b}{32\pi}}
  \begin{dcases}
    \left( \frac{R}{R_b} \right)^2 + \text{const.} ~,&R \le R_b \\
    4\sqrt{\frac{R}{R_b}} - 3 + \text{const.} ~,&R > R_b
    ,
  \end{dcases}
\end{align}
where $M_b = (4/3) \pi \ndrho_b R_b^3$ is the dimensionless reduced body mass. This result corresponds to the limit of small $r$ derived in Ref. \cite{SL10}. The choice of additive constant is arbitrary and may be chosen to match the full single-body solution.

Returning to Eq. (\ref{eqn:nd_vainshtein_pot}), we note that the coefficient $k$ corresponding to the two-body Sun-Earth system is in fact negligibly small. For example, with the Sun as the reference body and a reference distance $d=1$ AU, $k \approx 10^{-11}$. The distance 1 AU is well within the Vainshtein radii of the Sun, Earth, and Moon, which are approximately $3 \times 10^7$ AU, $4 \times 10^5$ AU, and $1 \times 10^5$ AU, respectively. For comparison, the apogee of Pluto's orbit is around 50 AU and the Oort cloud extends to at most $2 \times 10^5$ AU from the Sun. Furthermore, comparison between the analytic one-body solutions for the force field caused by the Sun, $\partial_r\ndphi_\textrm{S}$, for $k=0$ and $k \ne 0$ reveals that the linear term is indeed irrelevant in the Sun-Earth system, as the relative difference at 1 AU is only of the order of $10^{-11}$.

Smaller masses such as satellites or individual atoms have much smaller Vainshtein radii. However, so long as they are within the solar system, their potential fields will be dominated by the Sun or other planets at short distances below the Vainshtein radii of the smaller objects. For example, a hydrogen atom has a Vainshtein radius of approximately $0.4$ m, but the fifth force it would exert on an object one angstrom away is still tens of orders of magnitude smaller than the fifth force exerted by the Sun on an object at a distance of 1 AU. Similarly, a spherical satellite of mass $10^4$ kg has a Vainshtein radius of approximately $10^{10}$ m. However, at a distance of 1 AU from the Sun, the force the satellite would exert on a nearby mass is comparable to the fifth force exerted by the Sun at a millionth of an angstrom away. The magnitude of the Laplacian of its scalar field is comparable to that of the Sun at a distance of about 250 meters, still many orders of magnitude below the satellite's Vainshtein radius. In the present work, focusing on solar system scales, the coefficient $k$ of the linear term in Eq. (\ref{eqn:nd_vainshtein_pot}) was set to $0$ in all simulations, although the numerical method should remain valid for arbitrary $k>0$.

\section{Scalar potential solution for the axisymmetric Sun-Earth and 3D Sun-Earth-Moon systems}
\label{sec:results}
Although Eqs. (\ref{eqn:nd_vainshtein_pot}) and (\ref{eqn:vainshtein_nonlin_only}) have been solved analytically for the case of single body with spherical symmetry, as shown above, exact analytic solutions for asymmetric systems consisting of two or more bodies have remained intractable. By noting that the governing equation remains invariant to the addition of terms represented by constant gradients to $\ndphi$ (so-called Galilean invariance), Hui \textit{et al.} \cite{HS09} suggested that the influence of a distant mass on a local system could be approximated by first solving for the local system in isolation and then adding the linearized potential of the field of the distant mass. For an axisymmetric two-body problem, an analytic perturbation expansion based on this assumption has been developed \cite{C20}. To achieve higher accuracy for the two-body problem or to solve complex systems containing many bodies or non-spherical masses, it becomes necessary to turn to numerical solution techniques.

The success of the numerical method used in this work relies on an important observation by Chan and Scoccimarro \cite{CS09}. Recasting Eq. (\ref{eqn:nd_vainshtein_pot}) or Eq. (\ref{eqn:vainshtein_nonlin_only}) as a quadratic equation in terms of the Laplacian $\nabla^2 \ndphi$ allows one to isolate the solution which correctly matches the large-scale limiting behavior by selecting the corresponding positive or negative square root. While they used a discriminant splitting technique to avoid complex roots in the residual function of trial solutions in large-scale cosmological simulations containing both over- and under-densities, we find that the method without splitting is particularly useful for simulations like ours at distances below the Vainshtein radius containing dense and compact mass sources, i.e., cases in which underdensities can be ignored. In particular, we show that in this small scale regime, the residual error landscape of the solved quadratic form of Eq. (\ref{eqn:nd_vainshtein_pot}) or Eq. (\ref{eqn:vainshtein_nonlin_only}) has no local minima, implying that an iteration scheme following gradient descent will locally converge to the global minimum representing the true solution. The main analytic aspects of the iteration scheme used in the numerical simulations are discussed in Section \ref{sec:results:physical_basis}, while a detailed explanation of the implementation and numerical method of central finite differences with nested meshes is contained in Appendix \ref{sec:numerical_method}.

Unlike previous studies incorporating compact mass sources \cite{HS13,BH13}, the Vainshtein radii in our studies are many orders of magnitude larger than the radii and separation distance of the solar system bodies of interest. The scalar potential field is therefore computed well within the Vainshtein radii of the dominant bodies, without having to extend the computational domain to the far field region dominated by the linear term.
In addition, the boundary conditions applied along the edges of the computational domain derive from the values of the spherically-symmetric solution given by Eq. (\ref{eqn:ndOnebodySoln}) forced by a spherical average of the mass sources, in contrast to boundary conditions corresponding to superposition of single-body solutions.
The reader will find in Section \ref{sec:results:bcs} a more detailed explanation of the boundary conditions and validation tests are presented in Appendix \ref{sec:numerval:bcs}.

Experimental detection relying on force measurements would allow quantification of the Galileon force $\nabla \ndphi$, and its spatial variation in the form of the Laplacian $\nabla^2 \ndphi$. At solar system scales, $\nabla \ndphi$ is many orders of magnitude smaller than the force of Newtonian gravity. Directly measuring the small additional Galileon force would require exact computation of the gravitational field to the same precision, and is therefore one of the key obstacles to detection. However, because the Laplacian of the Newtonian gravitational field always vanishes, measurement of the Laplacian will reveal only non-Newtonian forces associated with the background scalar field. For this reason, we concentrate in this work mostly on the gradient and Laplacian functions of $\ndphi$ for the two-body Sun-Earth system and the three-body Sun-Earth-Moon system.

In Sections \ref{sec:results:se} and \ref{sec:results:sem}, we contrast the full numerical solutions for the 2D axisymmetric Sun-Earth and 3D Cartesian Sun-Earth-Moon system, with the solution to the single body Sun case, and na\"ive solutions based on simple superposition of the independent scalar fields. The results of the Sun-Earth-Moon system are further compared to the superposition of the two body Earth-Moon system with the single body Sun solution. Because the Sun's field has nearly a constant gradient in the region surrounding Earth and the Moon, the latter solution closely represents the approximation proposed by Hui \textit{et al.} \cite{HS09}. Simulation parameters are listed in Table \ref{table:simulationparams}. We note that although our numerical simulations were all based on the non-dimensionalized form of the governing equation and corresponding boundary conditions, the results that follow are presented in dimensional variables for the convenience of those readers interested in experimental scales and verification.

\begin{table*}
\setlength{\tabcolsep}{0pt}
\centering
\begin{tabular}{l rl rll}
\hline
\hline
&&&&& \\[-2pt]
~~~~~Quantity & \multicolumn{2}{c}{Scaling} & \multicolumn{3}{c}{Rescaled variable~~~~~~~}\\
\hline
&&&&& \\[-2pt]
\, Current Hubble rate constant \cite{CO17}& $H_0$ & $= 71 \,\textrm{km/s/Mpc} $ &&& \,
\\
\, Matter density parameter \cite{CO17} & $\Omega_m^0$ & $= 0.27 $ &&& \,
\\
\, Speed of light in vacuum& $c$ & $=2.998 \times 10^{8}$ m s$^{-1}$
\\
\, Crossover length scale& $r_c$ & $=c H_0^{-1}\left( 1 - \Omega_m^0 \right)^{-1}$ &&& \,
\\
\, & & $=1.8 \times 10^{23}$ km &&& \,
\\
\, Gravitational constant& $G$ & \multicolumn{4}{l}{$=6.674 \times 10^{-11}$ m$^3$ kg$^{-1}$ s$^{-2}$}
\\
&&&&& \\[-2pt]
\, Sun density& $\dimrho_{\text{S}}$ &$= 1,408~\text{kg m}^{-3}$ & $\ndrho_{\text{S}}$ &$= \dimrho_{\text{S}}/\dimrhoref$ &$= 1$ \,
\\
\, Sun radius& $r_{\text{S}}$ &$= 0.6957 \times 10^6$ km & $R_{\text{S}}$ &$= r_{\text{S}}/d$ &$= 1$ \,
\\
\, Sun Vainshtein radius& $r_{V,\text{S}}$ & \multicolumn{3}{l}{$= \left[ (64\pi/27) G c^{-2} \dimrho_{\text{S}} r_{\text{S}}^3 r_c^2 \right]^{1/3}$} & \,
\\
\, & &$= 4.396 \times 10^{15}$ km & $R_{V,\text{S}}$ &$= r_{V,\text{S}}/d$ &$= 6.318 \times 10^9$ \,
\\
\\[-2pt]
\, Sun coordinates~~& $x_{\text{S}}$ &$= 0$ & $X_{\text{S}}$ &$= x_{\text{S}}/d$ &$= 0$ \,
\\
\, & $y_{\text{S}}$ &$= 0$ & $Y_{\text{S}}$ &$= y_{\text{S}}/d$ &$= 0$ \,
\\
\, & $z_{\text{S}}$ &$= 0.5~\text{AU}$ & $Z_{\text{S}}$ &$= z_{\text{S}}/d$ &$= 107.5$ \,
\\
\, &  &$= 74.80 \times 10^6~\text{km}$ &  & & \,
\\
&&&&& \\[-2pt]
\, Earth density& $\dimrho_{\text{E}}$ &$= 5,515~\text{kg m}^{-3}$ & $\ndrho_{\text{E}}$ &$= \dimrho_{\text{E}}/\dimrhoref$ &$= 3.917$ \,
\\
\, Earth radius& $r_{\text{E}}$ &$= 0.006371 \times 10^6$ km & $R_{\text{E}}$ &$= r_{\text{E}}/d$ &$= 9.158 \times 10^{-3}$ \,
\\
\, Earth Vainshtein radius& $r_{V,\text{E}}$ &$= 6.346 \times 10^{13}$ km & $R_{V,\text{E}}$ &$= r_{V,\text{E}}/d$ &$= 9.121 \times 10^7$ \,
\\
\\[-2pt]
\, Earth coordinates~~& $x_{\text{E}}$ &$= 0$ & $X_{\text{E}}$ &$= x_{\text{E}}/d$ &$= 0$ \,
\\
\, & $y_{\text{E}}$ &$= 0$ & $Y_{\text{E}}$ &$= y_{\text{E}}/d$ &$= 0$ \,
\\
\, & $z_{\text{E}}$ &$= -74.80 \times 10^6~\text{km}$ & $Z_{\text{E}}$ &$= z_{\text{E}}/d$ &$= -107.5$ \,
\\
&&&&& \\[-2pt]
\, Moon density& $\dimrho_{\text{M}}$ &$= 3,344~\text{kg m}^{-3}$ & $\ndrho_{\text{M}}$ &$= \dimrho_{\text{M}}/\dimrhoref$ &$= 2.375$ \,
\\
\, Moon radius& $r_{\text{M}}$ &$= 0.001737 \times 10^6$ km & $R_{\text{M}}$ &$= r_{\text{M}}/d$ &$= 2.497 \times 10^{-3}$ \,
\\
\, Moon Vainshtein radius& $r_{V,\text{M}}$ &$= 1.464 \times 10^{13}$ km & $R_{V,\text{M}}$ &$= r_{V,\text{M}}/d$ &$= 2.105 \times 10^7$ \,
\\
\\[-2pt]
\, Moon coordinates& $x_{\text{M}}$ &$= 0$ & $X_{\text{M}}$ &$= x_{\text{M}}/d$ &$= 0$ \,
\\
\, & $y_{\text{M}}$ &$= -0.3850 \times 10^{6}~\text{km}$ & $Y_{\text{M}}$ &$= y_{\text{M}}/d$ &$= -0.5534$ \,
\\
\, & $z_{\text{M}}$ &$= -74.80 \times 10^6~\text{km}$ & $Z_{\text{M}}$ &$= z_{\text{M}}/d$ &$= -107.5$ \,
\\
&&&&& \\[-2pt]
\, Reference distance& $d$ &$= 0.6957 \times 10^6$ km $=r_S$& \multicolumn{2}{l}{~~~$(X,Y,Z,R)$} &$=(x,y,z,r)/d$ \,
\\
\, Reference body density& $\dimrhoref$ &$= 1,408~\text{kg m}^{-3}=\dimrho_\textrm{S}$ & $\ndrho$ &$=\dimrho/\dimrhoref$ & \,
\\
\, Reference body radius& $\rref$ &$= 0.6957 \times 10^6$ km = $r_\textrm{S}$ & $\Rref$ &$= \rref/d$ &$= 1$ \,
\\
\, Reference scalar field value& $\dimphiref$ &$=\sqrt{\left( 8 \pi G \dimrhoref \right)c^2 d \rref^{3}/r_c^2}$ & & & \,
\\
\, & &$=1.2 \times 10^{-3}$ m$^2$ s$^{-2}$ & $\ndphi$ & $= \dimphi/\dimphiref$& \,
\\[2pt]
&&&&& \\[-2pt]
\, Linear coefficient of Eq. (\ref{eqn:nd_vainshtein_pot})& $k$ & $= \sqrt{(9 d^3 c^2)/(8 \pi G \dimrhoref \rref^3 r_c^2)}$
\\
\, & &$= 3.2 \times 10^{-15}$ &$k$&\multicolumn{2}{l}{$=0$ \text{in simulations}} \,
\\
\\[-2pt]
\, Gravitational potential field& $\dimpsiG$ \! & : \phantom{=}$\nabla^2 \dimpsiG=4 \pi G \dimrho(\vec{r})$ & $\ndpsi_{\text{G}}$ &$= \dimpsiG/\dimphiref$ & \,
\\
&&&&& \\[-2pt]
\, Quantity normalized by gravity&& $\| \cdot \|_G=\| \cdot \|/\|\nabla \dimpsiG\|$
\\
&&&&& \\[-2pt]
\, Arbitrary body density& $\dimrho_{\text{B}}$ & & $\ndrho_{\text{B}}$ &\multicolumn{2}{l}{$= (\dimrho_{\text{B}}/\dimrhoref) (d/r_\textrm{ref})^3$}
\\
\, Arbitrary body radius& $r_{\text{B}}$ & & $R_{\text{B}}$ &$= r_{\text{B}}/d$ & \,
\\

&&&&& \\[-2pt]
\, Computational domain size& $\ell$ &$= 64~\text{AU}$ & $L$ &$= \ell/d$ &$= 1.376 \times 10^{4}$ \,
\\
\, & &$= 9,574 \times 10^6~\text{km}$ & & & \,
\\
\\[-2pt]
\, 3D simulation bounds& $(x,y$ & \multicolumn{4}{l}{$,z)\in [-\ell,\ell] \times [-\ell,\ell] \times [-\ell,\ell]$}
\\[2pt]
\, 2D simulation bounds& $(r$ & \multicolumn{4}{l}{$,z)\in [0,\ell] \times [-\ell,\ell]$}
\\
\, Iteration number & $n$
\\
\hline
\hline
\end{tabular}
\caption{Parameter values and scalings for the numerical simulations (unless otherwise specified). In this work, the  reference distance $d$, reference body density $\rho_\textrm{ref}$, and reference body radius $r_\textrm{ref}$ were set equal to the Sun values, resulting in a reference scalar field, $\dimphi_\textrm{ref}$, based on the Sun.}
\label{table:simulationparams}
\end{table*}

\subsection{Solution scheme}
\label{sec:results:physical_basis}
The numerical solution scheme for obtaining the CGG scalar potential field at solar system scales is based on inclusion of mass sources far denser than any local cosmological underdensities. Under this assumption, the mass density term $\ndrho$ of Eq. (\ref{eqn:nd_vainshtein_pot}) can be assumed to be non-negative, which allows formulation of a robust iteration scheme with rapid convergence regardless of the initial trial solution. The accuracy and convergence of this iteration scheme are examined next.

\subsubsection{Analytic properties of implemented iteration scheme}
\label{sec:results:iteration_scheme}
The solution to the general governing nonlinear equation given by Eq. (\ref{eqn:nd_vainshtein_pot}) can be accurately approximated by iterative linearization. Given a nonlinear residual function $\scR[\ndphi]$ quantifying the difference of an interim solution from the actual solution $\ndphi$, the numerical approximation scheme is recast as an optimization problem by minimizing the value of the integrated residual over the volume of interest, namely $\|\scR [\ndphi]\|^2 = \int \scR^2 [\ndphi] dV$. The initial trial function for $\ndphi$ is then made to evolve via gradient descent toward a minimum of the residual, where the gradient operator is defined by the functional derivative $\scL[\ndphi] = \delta \scR[\ndphi]/\delta \ndphi$. A variety of algorithms exist in the literature for speeding the computations involving gradient descent and seeking global minima amidst a residual landscape potentially populated by many local minima, all the while ensuring accuracy and stability \cite{K95}.

The choice of residual function is not unique and ultimately establishes the details of the residual landscape, which can complicate identification of the global minimum. The most straightforward option based simply on collection of all terms in Eq. (\ref{eqn:nd_vainshtein_pot}) yields a direct residual function $\scRd$ and direct linear gradient operator $\scLd$ given by
\begin{align}
  \scRd[\ndphi] &=
  k \nabla^2 \ndphi + \left[ \left( \nabla^2 \ndphi \right)^2 - \sum_{i,j} \left(\nabla_{i}\nabla_{j} \ndphi\right)^2 \right] - \ndrho
  ,
  \\
  \scLd[\ndphi] &= k \nabla^2 + 2 \left( \nabla^2 \ndphi \right) \nabla^2 - 2 \sum_{i,j} \left( \nabla_i \nabla_j \ndphi \right) \nabla_i \nabla_j
  .
\end{align}
This choice of residual and linear operator has previously been shown to produce convergence in cases where the initial trial function was chosen to be close to the true solution \cite{HS13,BH13,OK19} or in cases where the governing equation was restricted to the large-scale regime where the linear term is dominant \cite{Sch09a}. The difficulty in applying this choice of residual function to finding solutions of Eq. (\ref{eqn:nd_vainshtein_pot}) is that its quadratic form can yield two solution branches, leading to a residual landscape containing at least two global minima [we say ``at least,'' because proof that there exist only two solutions to Eq. (\ref{eqn:nd_vainshtein_pot}) would require analysis beyond the scope of this paper]. Furthermore, our numerical tests have found that if the trial solution is not sufficiently close to the true solution, then gradient descent with this direct residual can yield solutions which settle into minima far from the true solution. Such local minima can occur when the local solution in one region of space iterates toward the repulsive branch of Eq. (\ref{eqn:nd_vainshtein_pot}) while the solution in a different region iterates towards the attractive branch.

Chan and Scoccimarro \cite{CS09} made the critical observation that upon solving Eq. (\ref{eqn:nd_vainshtein_pot}) as a quadratic equation in $\nabla^2 \ndphi$, one can explicitly select a solution branch and thus avoid the potential problem of different points converging to the undesired branch when the starting trial solution is not close enough to the true solution.
With that insight, the positive branch is given by 
\begin{align}
  \label{eqn:nd_vainshtein_pot_sq}
  \nabla^2 \ndphi &= \sqrt{\sum_{i,j} \left(\nabla_{i}\nabla_{j} \ndphi\right)^2 + \ndrho + \left( \frac{k}{2} \right)^{2}} - \frac{k}{2}
  .
\end{align}
This then leads to the following natural choice for the residual function and gradient operator:
\begin{align}
  \label{eqn:residual}
  \scR[\ndphi] &=
  \sqrt{\sum_{i,j} \left(\nabla_{i}\nabla_{j} \ndphi\right)^2 + \ndrho + \left( \frac{k}{2} \right)^{2}} - \nabla^2 \ndphi - \frac{k}{2}
  ,
  \\
  \label{eqn:linear_operator}
  \scL[\ndphi] &=
  \frac{\sum_{i,j} \left( \nabla_i \nabla_j \ndphi \right) \nabla_i \nabla_j}{\sqrt{\sum_{\ell,m} \left(\nabla_{\ell}\nabla_{m} \ndphi\right)^2 + \ndrho + \left( \frac{k}{2} \right)^{2}}} - \nabla^2
  .
\end{align}
For any solution $\ndphi$ of Eq. (\ref{eqn:nd_vainshtein_pot}), the discriminant in the square root will be positive. However, when evaluating the residual for a trial solution which is not a true solution, the discriminant will not necessarily be positive if $\ndrho < 0$. At galactic and cosmological scales, such $\ndrho < 0$ underdensities must be considered, so Chan and Scoccimarro used a discriminant splitting method to ensure that the residual could be evaluated for any trial solution. In the small-scale case around dense bodies, underdensities can be ignored, i.e., $\ndrho \ge 0$, and hence splitting the discriminant is unnecessary.

As we demonstrate next, this reformulation introduces a significant advantage for computation in that all critical points of $\scR[\ndphi]^2$ are global minima, due to the fact that $\scR[\ndphi]$ is a convex function unbounded from below. The convexity of $\scR[\ndphi]$ is evident from inspection of the second functional derivative of $\scR$ acting on an arbitrary function $\xi$, which is always non-negative:
\begin{align}
  \notag
  \frac{\delta^2\scR}{\delta \ndphi^2}[\xi,\xi] &
  = \frac{\left[\ndrho + \left( \frac{k}{2} \right)^2 + \sum_{i,j} \left( \nabla_i \nabla_j \ndphi \right)^2\right]\sum_{i,j}\left( \nabla_i \nabla_j \xi \right)^2}{\left[\sum_{\ell,m} \left(\nabla_{\ell}\nabla_{m} \ndphi\right)^2 + \ndrho + \left( k/2 \right)^{2}\right]^{3/2}}
  \\\notag&
  - \frac{\left[\sum_{i,j} \left( \nabla_i \nabla_j \ndphi \right) \left(\nabla_i \nabla_j \xi \right)\right]^2}{\left[\sum_{\ell,m} \left(\nabla_{\ell}\nabla_{m} \ndphi\right)^2 + \ndrho + \left( k/2 \right)^{2}\right]^{3/2}}
  \\&
  \ge \frac{\left[\ndrho + \left( \frac{k}{2} \right)^2 \right]\left[ \sum_{i,j}\left( \nabla_i \nabla_j \xi \right)^2\right]}{\left[\sum_{\ell,m} \left(\nabla_{\ell}\nabla_{m} \ndphi\right)^2 + \ndrho + \left( k/2 \right)^{2}\right]^{3/2}}
  \ge 0
  .
\end{align}
The functional second derivative vanishes only when $\xi$ is a constant or a linear function. Both forms of $\xi$ represent a gauge freedom of $\ndphi$ since any constant or linear function can be added to a solution of Eq. (\ref{eqn:nd_vainshtein_pot}) and remain a solution; hence, $\scR$ is convex. The unboundedness of $\scR$ from below can be shown by considering the ansatz function $\ndphi = (c/2)(x^2 + y^2 + z^2)$ representing solutions close to an extremum. Then $\nabla^2 \ndphi = 3c$ and $\sum_{i,j}(\nabla_i \nabla_j \ndphi)^2 = 3 c^2$. If $c \gg k$ and $c \gg \ndrho$, then $\scR[\ndphi] \approx (\sqrt{3} -3)c$, which can assume arbitrarily large values for arbitrarily large coefficient values $c$.

Because $\scR$ is a convex function unbounded from below, $\scR^2$ has the property that the only critical points are global minima. This can be seen immediately by noting that $\delta \scR^2/\delta \ndphi = 2 \scR (\delta \scR/\delta \ndphi)$, which can vanish only if $\delta \scR/\delta \ndphi = 0$ or $\scR = 0$.
But the existence of an extremum satisfying $\delta \scR/\delta \ndphi = 0$ would contradict the unboundedness of $\scR$, and hence the only critical points of $\scR^2$ correspond to points at which $\scR=0$, which represent global minima of $\scR^2$.
Furthermore, this implies that so long as $\scR$ is sufficiently smooth, the global minima of $\scR^2$ must be connected, in the sense that one solution can be continuously deformed into another while satisfying the global minimum condition $\scR^2 = 0$. Were there to exist two separated minima, there would then have to exist a non-minimum critical point on a line connecting them, resulting in a contradiction.

The properties of $\scR$ and $\scR^2$ so far described represent local behavior. However, the residual function of interest, which represents a global constraint, is represented by the integral $L_2$ norm of $\scR$, namely $\|\scR[\ndphi]\|^2 = \int \scR^2[\ndphi] dV$. Because the global minima of $\scR^2$ are connected within  sufficiently smooth regions of $\scR$, then if boundary conditions allow a global solution to exist, the quantity $\|\scR[\ndphi]\|^2$ is also expected to have no minima aside from the global minimum.
Rigorous proof is beyond the scope of this paper, as is proof of the existence and uniqueness of solutions to $\scR[\ndphi]=0$ subject to Dirichlet boundary conditions. Furthermore, the strong concavity of the landscape requires $k>0$; if $k$ is taken to be 0 then there are no local minima but there may be saddle points. That said, we have found that in practice the numerical convergence of $\|\scR[\ndphi]\|^2$ is accurate, stable, and rapidly convergent even in the $k=0$ limit, suggesting that the local properties of $\scR^2$ yield a residual landscape for $\|\scR[\ndphi]\|^2$, whose geometry is highly favorable to gradient descent techniques and rapid identification of the global minimum.

\subsubsection{Boundary conditions in numerical simulations}
\label{sec:results:bcs}
Solution of Eq. (\ref{eqn:nd_vainshtein_pot_sq}), which is second-order, requires specification of a condition at each point on the boundary. For the axisymmetric Sun-Earth simulations, we invoked a Neumann condition reflecting  symmetry about the $R=0$ axis such that $\partial_R \ndphi(R=0,Z) = 0$. With regard to the remaining far field boundary conditions in $R$ and $Z$, (or in $X$, $Y$, and $Z$, for the 3D Sun-Earth-Moon simulations based on Cartesian geometry), we note the following reasoning for the choice of Dirichlet conditions.

When simulations of Eq. (\ref{eqn:nd_vainshtein_pot_sq}) are conducted in a computational domain whose size is much larger than the Vainshtein radii of the interior bodies, the equation becomes dominated by the linear term along the far field exterior boundaries. In this case, boundary conditions based on superposition of the individual analytic single-body solutions may represent a good choice \cite{HS13}. In the present study, however, the Vainshtein radii are prohibitively large and all simulations were conducted within a computational domain whose size represents relatively small scales such that the nonlinear term in Eq. (\ref{eqn:nd_vainshtein_pot_sq}) is dominant.  For such a nonlinear equation, there is no reason to expect that the boundary conditions applied along the domain boundaries should be accurately represented by simple superposition of single-body solutions. However, it is expected that so long as the domain edges are sufficiently far from the included bodies, they should together act as a point source or equivalently, the scalar potential function should behave as though it is driven by a single point mass.

Since in our simulations all bodies were confined to the interior of the computational domain, we adopted far field Dirichlet conditions obtained from the value of the scalar potential given by Eq. (\ref{eqn:spherk0}) for a point mass equal to the total mass of all interior bodies positioned at the center of mass of those bodies. Were the computational domain to be spherical, this boundary condition would be a constant applied on the domain boundaries. But because the computational domain was either spherical or cubic, the spherically-symmetric solution was used to determine the values at each point of the boundary, resulting in a non-constant boundary condition. In what follows, we refer to this choice of boundary condition as the point source boundary condition (PSBC) and its dimensional value denoted by $\dimphi_\infty(\vec{r})$ [or dimensionless value $\Phi_\infty(\vec{R})$]. It should also be  noted that for a spherically-symmetric system whose density field has compact support, the scalar potential field in the external vacuum depends only on the total mass and not its spatial distribution. Thus, a point mass and an arbitrary compact spherically-symmetric mass distribution are indistinguishable beyond their radii, and the point mass boundary condition is equivalent to the solution of the scalar potential equation forced by a spherical average of the density field. The point source boundary condition is therefore the natural physical choice for the scalar potential field at distances much greater than the separation distances of the interior bodies. For the solar system, the Sun is so massive that the relative difference between the point source solution and the linear superposition of single-body solutions is of the order of $10^{-6}$ and therefore essentially negligible. However, it seems inappropriate to impose far field boundary conditions based on linear superposition of individual single-body potential fields when solving a nonlinear equation.

\subsection{Results of axisymmetric Sun-Earth system}
\label{sec:results:se}
\begin{figure*}[!htb]
\centering
\includegraphics[scale=0.8]{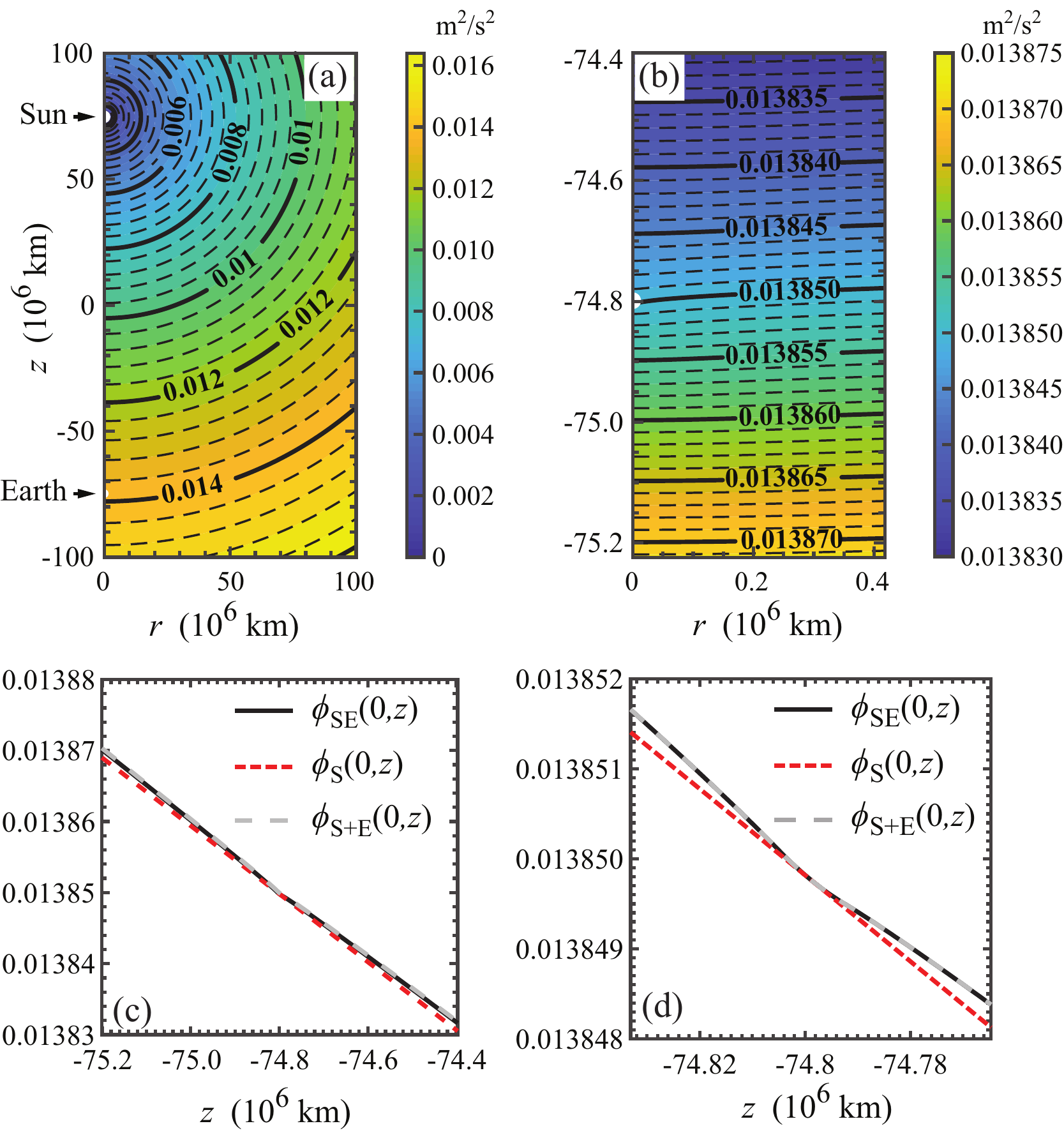}
\caption{
\label{fig:sun_earth_plots_phi}
Numerical solutions for the axisymmetric Sun-Earth (SE) Galileon potential field $\dimphiASE (r,z)$ $\text{m}^2/\text{s}^2$. Magnitudes indicated on solid contour lines (black) correspond to major divisions on color bar; dashed contour lines represent 1/5 intermediate color bar values. Sun and Earth bodies shown in white. (a) Contour plot for region containing Sun (S) and Earth (E) bodies positioned at coordinate values ($r_{\textrm{S}}=0, z_{\textrm{S}}= + 74.80 \times 10^6 \textrm{km} = + 0.5 \, \textrm{AU})$ and ($r_{\textrm{E}}=0, z_{\textrm{E}}= -74.80 \times 10^6 \textrm{km} = - 0.5 \,\textrm{AU})$.  (b) Magnified view of (a) showing solution about the Earth center. (c) Comparison of three solutions in the vicinity of the Earth body along the line connecting the Sun and Earth bodies: full solution $\dimphiASE$ (solid black line), single-body Sun solution $\dimphiS$ (dashed red line) and combined solution $\dimphiSpE$ (dashed gray line) from linear superposition of single-body Earth and Sun solutions. Span in $z$ equals a distance $4.17 \times 10^5$ km about Earth. (d) Magnified view of solutions in (c). Span in $z$ equals a distance $3.48 \times 10^4$ km about Earth.
}
\end{figure*}

Shown in Fig. \ref{fig:sun_earth_plots_phi} are far-field and near-field views about the Earth body of the dimensional axisymmetric Sun-Earth Galileon field $\dimphiASE (r,z)$ $\text{m}^2/\text{s}^2$ for Sun (S) and Earth (E) bodies positioned on the axis of symmetry $r=0$. The body coordinates were chosen to be ($r_{\textrm{S}}=0, z_{\textrm{S}}= + 74.80 \times 10^6 \,\textrm{km} = + 0.5 \,\textrm{AU})$ and ($r_{\textrm{E}}=0, z_{\textrm{E}}= -74.80 \times 10^6 \,\textrm{km} = - 0.5 \, \textrm{AU})$. The boundary conditions applied along the exterior edges of the cylindrical domain were $\partial_r \dimphiASE(r=0, -64 \,\textrm{AU} \leq z \leq +64 \, \textrm{AU})=0$, $\dimphiASE(r=+64 \,\textrm{AU}, -64 \, \textrm{AU} \leq z \leq +64 \, \textrm{AU})= \dimphi_\infty(r,z)$ and $\dimphiASE(0 \leq r \leq 64 \, \textrm{AU}, z = \pm 64 \, \textrm{AU}) = \dimphi_\infty(r,z)$. The strong spherical symmetry of the solution about the Sun body evident in Fig. \ref{fig:sun_earth_plots_phi}(a) is indicative of the fact that the field is dominated by the massive Sun. Shown in Fig. \ref{fig:sun_earth_plots_phi}(b) is a magnified view of the field about the Earth body. The results in Fig. \ref{fig:sun_earth_plots_phi}(c) and the magnified view in (d) depict the field values along the axis of symmetry near the Earth body. The results show a slight reduction in the field value near the  location of the Earth body. Shown for comparison is the single-body Sun solution $\dimphiS(0,z)$ and the combined  solution from linear superposition of the single-body Sun and Earth solutions $\dimphiSpE(0,z)$. At the scales shown about the Earth body, the full numerical solution $\dimphiASE(0,z)$ and the solution obtained by linear superposition of single-body solutions $\dimphiSpE(0,z)$ are virtually indistinguishable but differ from the single-body Sun solution $\dimphiS(0,z)$. Whereas $\dimphiS(0,z)$ appears nearly linear throughout the range shown, the full solution given by $\dimphiASE$ contains a visible bend within a distance of approximately $O(10^5~\text{km})$ of the Earth center.

\begin{figure*}[!htb]
\centering
\includegraphics[scale=0.8]{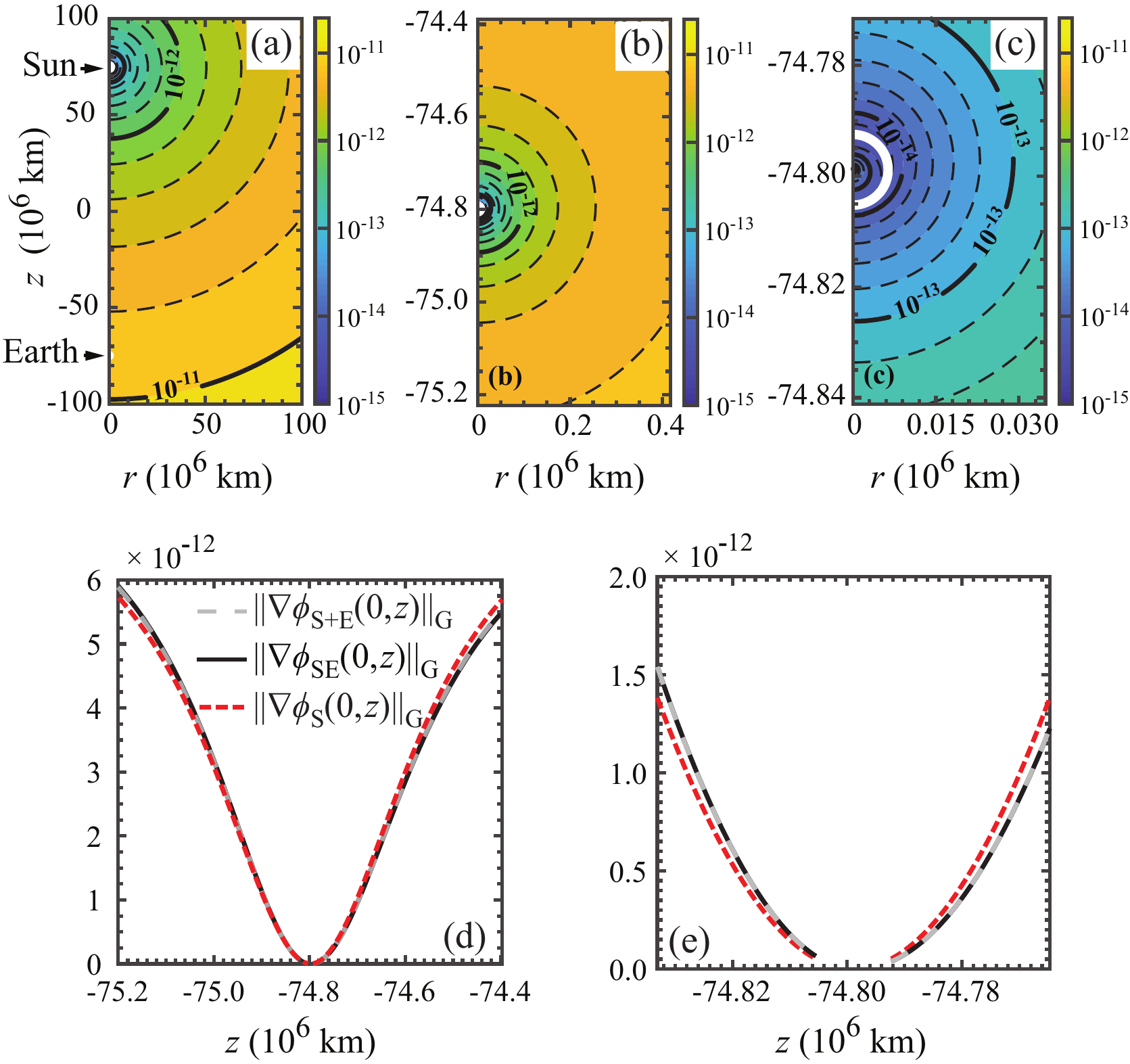}
\caption{
\label{fig:sun_earth_forcestrength}
Numerical solutions for the axisymmetric Sun-Earth (SE) force field normalized by the force of gravity (G), $\|\nabla \dimphiASE(r,z)\|_G = \|\nabla \dimphiASE(r,z)\|/\|\nabla \dimpsiG(r,z)\|$, where $\dimpsiG(r,z)$ is the Newtonian gravitational potential. Magnitudes plotted on a logarithmic scale and indicated by solid contour lines (black) correspond to major divisions on color bar; dashed contour lines represent 1/5 intermediate color bar values. Sun and Earth bodies shown in white. (a) Contour plot in region containing Sun (S) and Earth (E) bodies positioned at coordinate values ($r_{\textrm{S}}=0, z_{\textrm{S}}= + 74.80 \times 10^6 \textrm{km} = + 0.5 \, \textrm{AU})$ and ($r_{\textrm{E}}=0, z_{\textrm{E}}= -74.80 \times 10^6 \textrm{km} = - 0.5 \, \textrm{AU})$. Magnitudes plotted on a logarithmic scale. (b) Magnified view of solution in (a) centered about the Earth body. Span in $z$ is 10\% larger than the Moon's orbit radius. (c) Further magnified view of contour plot in (b). Earth body outlined in white. (d) Comparison of three solutions in the vicinity of the Earth body along the line connecting the Sun and Earth bodies: full solution $\|\nabla \dimphiASE\|_G$ (solid black line), single-body Sun solution  $\|\nabla \dimphiS\|_G$ (dashed red line) and combined solution $\|\nabla \dimphiSpE\|_G$ (dashed gray line) from linear superposition of the single-body Earth and Sun solutions. Span in $z$ equals a distance $4.17 \times 10^5$ km about Earth. (e) Magnified view of solutions in (d). Span in $z$ equals a distance $3.48 \times 10^4$ km about Earth.
}
\end{figure*}

Shown in Fig. \ref{fig:sun_earth_forcestrength} is a large scale view and near field views about the Earth body of the relative strength of the fifth force to the force of Newtonian gravity, $\|\nabla \dimphiASE(r,z)\|_G = \|\nabla \dimphiASE(r,z)\| / \|\nabla \dimpsiG\|$, where $\|\cdot\|$ denotes the vector norm. These data correspond to the simulation runs shown in Fig. \ref{fig:sun_earth_plots_phi}. Here $\dimpsiG$, the Newtonian potential, is the solution of $\nabla^2\dimpsiG(\vec{r}) = 4 \pi G \rho(\vec{r})$. The strong spherical symmetry about the massive Sun body is evident in Fig. \ref{fig:sun_earth_forcestrength}(a). The magnified plots in Fig. \ref{fig:sun_earth_forcestrength}(b) and (c) also indicate high spherical symmetry about the Earth body with only slight elongation along the $z$ axis. The results in Fig. \ref{fig:sun_earth_forcestrength}(d) and the magnified view in (e) depict the spatial variation in the field along the axis of symmetry near the Earth body. The plots shown exclude results within the regions interior to the Earth body where the gravitational force vanishes. The results show a very slight depression near the location of the Earth body with a slight asymmetry about its center. Shown for comparison is the single-body Sun solution and the combined solution from superposition of the single-body Sun and Earth solutions. At the scales about the Earth body indicated, the full numerical solution and the superposed solution are virtually indistinguishable but differ from the single-body Sun solution. The visible asymmetry between the solutions reflects the fact that the Sun's and Earth's force fields oppose each other on the side of Earth facing the Sun and supplement each other on the side of Earth away from the Sun.

\begin{figure*}[!htb]
\centering
\includegraphics[scale=0.8]{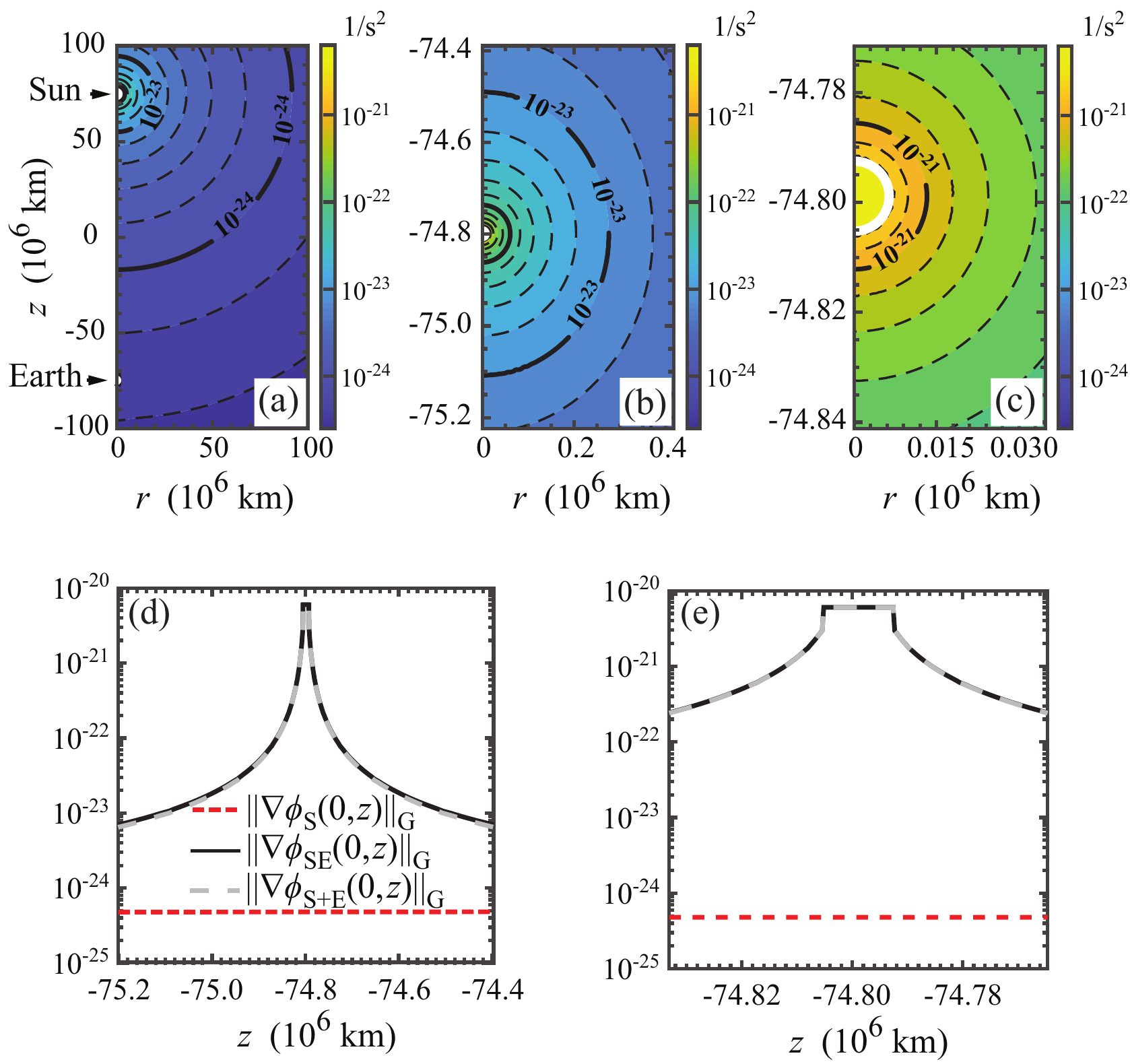}
\caption{
\label{fig:sun_earth_laplacian}
Contour plots showing the axisymmetric Sun-Earth (SE) Laplacian field distribution $\nabla^2 \dimphiASE(r,z)$ $\text{s}^{-2}$ plotted on a logarithmic scale. Magnitudes plotted on a logarithmic scale and indicated by solid contour lines (black) correspond to major divisions on color bar; dashed contour lines represent 1/5 intermediate color bar values. Sun and Earth bodies shown in white. (a) Solution in region containing both Sun (S) and Earth (E) bodies positioned at coordinate values ($r_{\textrm{S}}=0, z_{\textrm{S}}= + 74.80 \times 10^6 \textrm{km} = + 0.5 \, \textrm{AU})$ and ($r_{\textrm{E}}=0, z_{\textrm{E}}= -74.80 \times 10^6 \textrm{km} = - 0.5 \, \textrm{AU})$. (b) Magnified view of solution in (a) centered about the Earth body. Span in $z$ is 10\% larger than the Moon's orbit radius. (c) Magnified view of solution in (b). Earth body outlined in white. (d) Comparison of three solutions in the vicinity of the Earth body along the line connecting the Sun and Earth bodies: full solution $\nabla^2 \dimphiASE$ (solid black line), single-body Sun solution $\nabla^2 \dimphiS$ (dashed red line) and combined solution $\nabla^2 \dimphiSpE$ (dashed gray line) from linear superposition of the single-body Earth and Sun solutions. Span in $z$ equals a distance $4.17 \times 10^5$ km about Earth. (e) Magnified view of solutions in (d). Span in $z$ equals a distance $3.48 \times 10^4$ km about Earth.
}
\end{figure*}

Shown in Fig. \ref{fig:sun_earth_laplacian} are the results for the dimensional axisymmetric Sun-Earth (SE) Laplacian field $\nabla^2 \dimphiASE(r,z)$ $\text{s}^{-2}$ plotted on a logarithmic scale for the runs shown in Fig. \ref{fig:sun_earth_plots_phi}. The strong spherical symmetry of the solution about the massive Sun body is evident in Fig. \ref{fig:sun_earth_laplacian}(a). The magnified plots in Fig. \ref{fig:sun_earth_laplacian}(b) and (c) make evident the anisotropy along the $z$ axis due to the Sun body. The Laplacian field magnitude undergoes rapid decay with increasing distance from either body.  The results in Fig. \ref{fig:sun_earth_plots_phi}(d) and the magnified view in (e) depict the Laplacian field values along the axis of symmetry near the Earth body. Shown for comparison is the single-body Sun solution and the superposed single-body Sun and Earth solutions. At the scales about the Earth body indicated, the full numerical solution and the superposed solution are virtually indistinguishable but differ significantly from the single-body Sun solution in form and magnitude. In particular, the Laplacian field of the single-body Sun solution is uniformly negligible by comparison. Also evident from Fig. \ref{fig:sun_earth_plots_phi}(d) is the fact that the solution obtained from superposition everywhere slightly underestimates the correct magnitude, with the discrepancy increasing with distance from the Earth body. Fig. \ref{fig:sun_earth_plots_phi}(e) shows that the Laplacian field for the Sun-Earth system within the radius of the Earth is approximately constant, beyond which it undergoes rapid decay in accord with the single-body solution given by Eq. (\ref{eqn:spherk0}).

\begin{figure*}[!htb]
\centering
\includegraphics[scale=0.8]{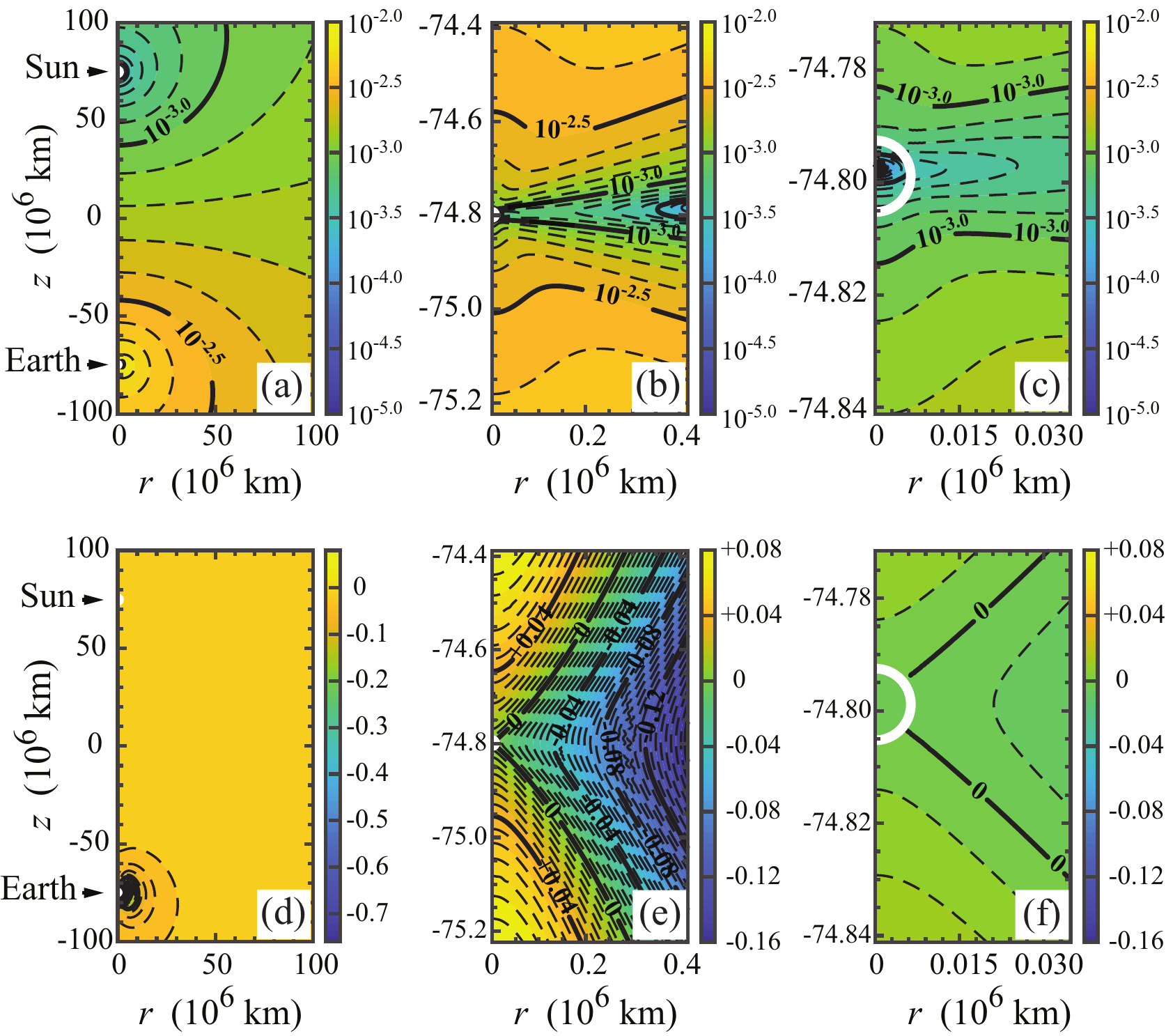}
\caption{
\label{fig:sun_earth_plots_diffs}
Contour plots showing the normalized differences $\|\nabla \dimphiASE(r,z) - \nabla \dimphiSpE(r,z)\|/\|\nabla \dimphiASE(r,z)\|$ (top panel) and $[\nabla^2 \dimphiASE(r,z) - \nabla^2\dimphiSpE(r,z)]/[\nabla^2 \dimphiASE(r,z)]$ (bottom panel) plotted on a logarithmic scale. Magnitudes indicated by solid contour lines (black) (logarithmic scale in top panel; linear scale on bottom panel) correspond to major divisions on corresponding color bar; dashed contour lines represent 1/5 intermediate color bar values. Sun and Earth bodies shown in white. (a) and (d) Solutions in region containing Sun (S) and Earth (E) bodies positioned at the coordinate values ($r_{\textrm{S}}=0, z_{\textrm{S}}= + 74.80 \times 10^6 \textrm{km} = + 0.5 \, \textrm{AU})$ and ($r_{\textrm{E}}=0, z_{\textrm{E}}= -74.80 \times 10^6 \textrm{km} = - 0.5 \, \textrm{AU})$. (b) and (e) Magnified view of solutions in (a) and (d) centered about the Earth body. Span in $r$ and $z$ is 10\% larger than the Moon's orbit radius. (c) and (f) Magnified view of solutions in (b) and (d) in close vicinity of the Earth body. Span in $r$ equals a distance $3.48 \times 10^4$ km about Earth.
}
\end{figure*}

Shown in Fig. \ref{fig:sun_earth_plots_diffs} are numerical solutions of the normalized differences for the Galileon force and Laplacian fields, namely $\|\nabla \dimphiASE(r,z) - \nabla \dimphiSpE(r,z)\|/\|\nabla \dimphiASE(r,z)\|$ (top panel) and $[ \nabla^2 \dimphiASE(r,z) - \nabla^2\dimphiSpE(r,z)]/[\nabla^2 \dimphiASE(r,z)]$ (bottom panel), plotted on a logarithmic scale. The relative errors are smaller near the Sun body than the Earth body. As evident from Figs. \ref{fig:sun_earth_plots_diffs}(a) and (d), these smaller errors in the vicinity of the Earth are caused by the fact that the more massive Sun body has a relatively larger influence on the field about the Earth than vice versa. As evident also from Figs. \ref{fig:sun_earth_plots_diffs}(c) and (f), for distances close to the Earth body, the relative error in the force field is of the order of $0.1$\% while that for the Laplacian field is of the order of $1$\%. At a distance of $4 \times 10^5$ km from Earth, these differences become larger - the relative error in the Laplacian field can exceed $15$\%, as shown in (e). In general too, the superposition approximation tends to underestimate the value of the Laplacian field along the central Sun-Earth axis and to overestimate the value away from this axis.

\subsection{Results of Sun-Earth-Moon system}
\label{sec:results:sem}
In this section we review results of 3D simulations for the three-body Sun-Earth-Moon system computed in a cubic domain (Cartesian coordinates). The Sun (S) and Earth (E) bodies were positioned on the $z$ axis and the Moon (M) located at a point in its orbit forming a $90^\circ$ angle with the Earth and Sun. The actual coordinates used in the simulations were as follows: ($x_{\textrm{S}}=0, y_{\textrm{S}}=0, z_{\textrm{S}}= + 74.80 \times 10^6 \, \textrm{km} = + 0.5 \, \textrm{AU})$, ($x_{\textrm{E}}=0, y_{\textrm{E}}=0, z_{\textrm{E}}= - 74.80 \times 10^6 \, \textrm{km} = - 0.5 \, \textrm{AU})$ and ($x_{\textrm{M}}=0, y_{\textrm{M}}= - 0.3845  \times 10^6 \, \textrm{km} = - 0.00257 \, \textrm{AU}, z_{\textrm{M}}= - 74.80 \times 10^6 \, \textrm{km} = - 0.5 \, \textrm{AU})$. The Dirichlet boundary conditions applied along the exterior edges of the cubic domain $-64 \, \textrm{AU} \leq (x,y,z) \leq +64 \, \textrm{AU}$ equaled those values given by Eq. (\ref{eqn:spherk0}) for a point particle with a mass equal to the total mass of the three individual bodies positioned at the location of the three-body center of mass.

\begin{figure*}[!h]
\centering
\includegraphics[scale=0.8]{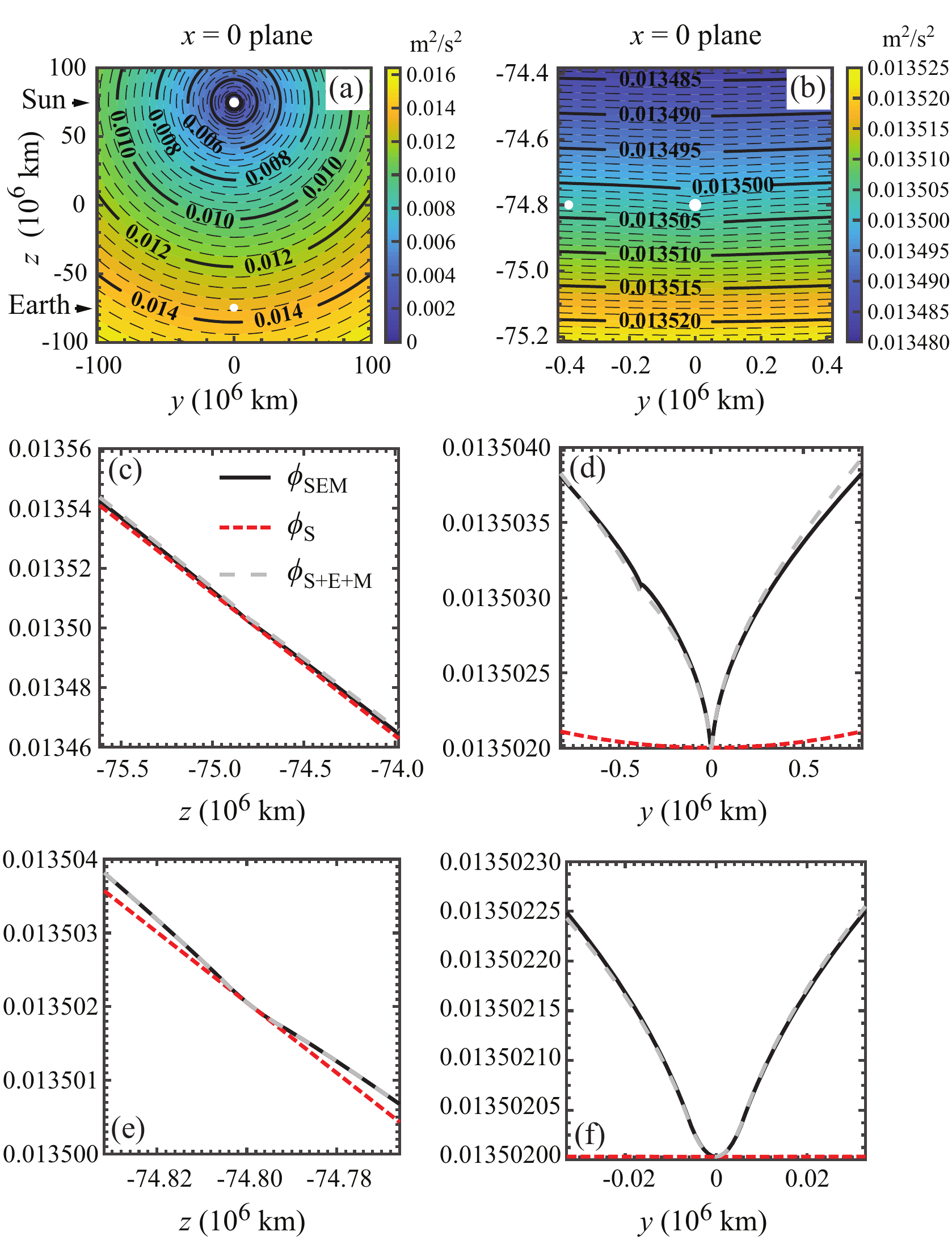}
\caption{
\label{fig:sun_earth_moon_plots_phi}
Numerical solutions for the Sun-Earth-Moon (SEM) scalar potential field $\dimphiSEM(x,y,z)$ [$\text{m}^2/\text{s}^2$] displayed in the  $x=0$ plane. Magnitudes indicated on solid contour lines (black) correspond to major divisions on color bar; dashed contour lines represent 1/5 intermediate color bar values. Sun, Earth and Moon bodies shown in white. (a) Contour plot in region containing Sun (S), Earth (E) and Moon (M) bodies positioned at the coordinate values ($x_{\textrm{S}}=0, y_{\textrm{S}}=0, z_{\textrm{S}}= + 74.80 \times 10^6 \, \textrm{km} = + 0.5 \, \textrm{AU})$, ($x_{\textrm{E}}=0, y_{\textrm{E}}=0, z_{\textrm{E}}= - 74.80 \times 10^6 \, \textrm{km} = - 0.5 \, \textrm{AU})$ and ($x_{\textrm{M}}=0, y_{\textrm{M}}= - 0.3845  \times 10^6 \, \textrm{km} = - 0.00257 \, \textrm{AU}, z_{\textrm{M}}= - 74.80 \times 10^6 \, \textrm{km} = - 0.5 \, \textrm{AU})$. (b) Magnified view of solution in (a) centered about the Earth body with the Moon to its left. (c) Comparison of three solutions in the vicinity of the Earth body along the line connecting the Sun and Earth bodies: full solution $\dimphiSEM$ (solid black), single-body Sun solution $\dimphiS$ (dashed red) and combined solution $\dimphiSpEpM$ (dashed gray) from linear superposition of the single-body Earth, Sun and Moon solutions. Span in $z$ equals a distance $8.35 \times 10^5$ km about Earth. (d) Comparison of three solutions in the vicinity of the Earth body along the line connecting the Earth and Moon: full solution $\dimphiSEM$, single-body Sun solution $\dimphiSEM$ and combined solution $\dimphiSpEpM$ from linear superposition of the single-body Earth, Sun and Moon solutions. (e) Magnified view of solutions in (c) in the vicinity of the Earth body. (f) Magnified view of solutions in (d) in the vicinity of the Earth body.
}
\end{figure*}

Shown in Fig.\ref{fig:sun_earth_moon_plots_phi} are the numerical solutions for the Sun-Earth-Moon (SEM) scalar potential field $\dimphiSEM(x,y,z)$ $\text{m}^2/\text{s}^2$. The strong spherical symmetry of the solution about the Sun body is evident in Fig. \ref{fig:sun_earth_moon_plots_phi}(a), indicative of the fact that the potential field is dominated by that of the massive Sun. Shown in Fig. \ref{fig:sun_earth_moon_plots_phi}(b) is a magnified view of the potential field near the Earth and Moon showing how their presence slightly distorts the local potential field. The plots in (c) and (e) show magnified views at distances close to the Earth body for the  potential field along the line joining the Sun and Earth. There is no visible difference between the plots in (c) and (e) and the values plotted in Fig. \ref{fig:sun_earth_plots_phi} (c) and (d), except for the addition of an overall constant which has no effect on the force. The results in (c) and (e) also show a slight reduction  in the field near the Earth body. Shown for comparison is the single-body Sun and combined solution from superposition of the single-body Sun, Earth and Moon solutions. At the scales shown about the Earth body, the full numerical solution and the solution obtained by linear superposition are quite close but differ from the single-body Sun solution. While the single-body solution exhibits linear behavior, the full solution contains a visible bend centered about the Earth body. The results in (d) and (f) clearly show the influence of the Moon on the potential field solution in close proximity to the Earth. Here, the deviations of the full and linear superposition solutions from the single-body Sun solution are more evident. In particular, the influence of the Moon is clearly visible by the kink appearing on the curve at $y=-0.385 \times 10^6$ km. Close inspection also reveals that the full solution differs somewhat from the superposition solution. The latter appears to underestimate the correct field value near the Moon and to overestimate the value on the side of the Earth farthest from the Moon.

\begin{figure*}[!h]
\centering
\includegraphics[scale=0.8]{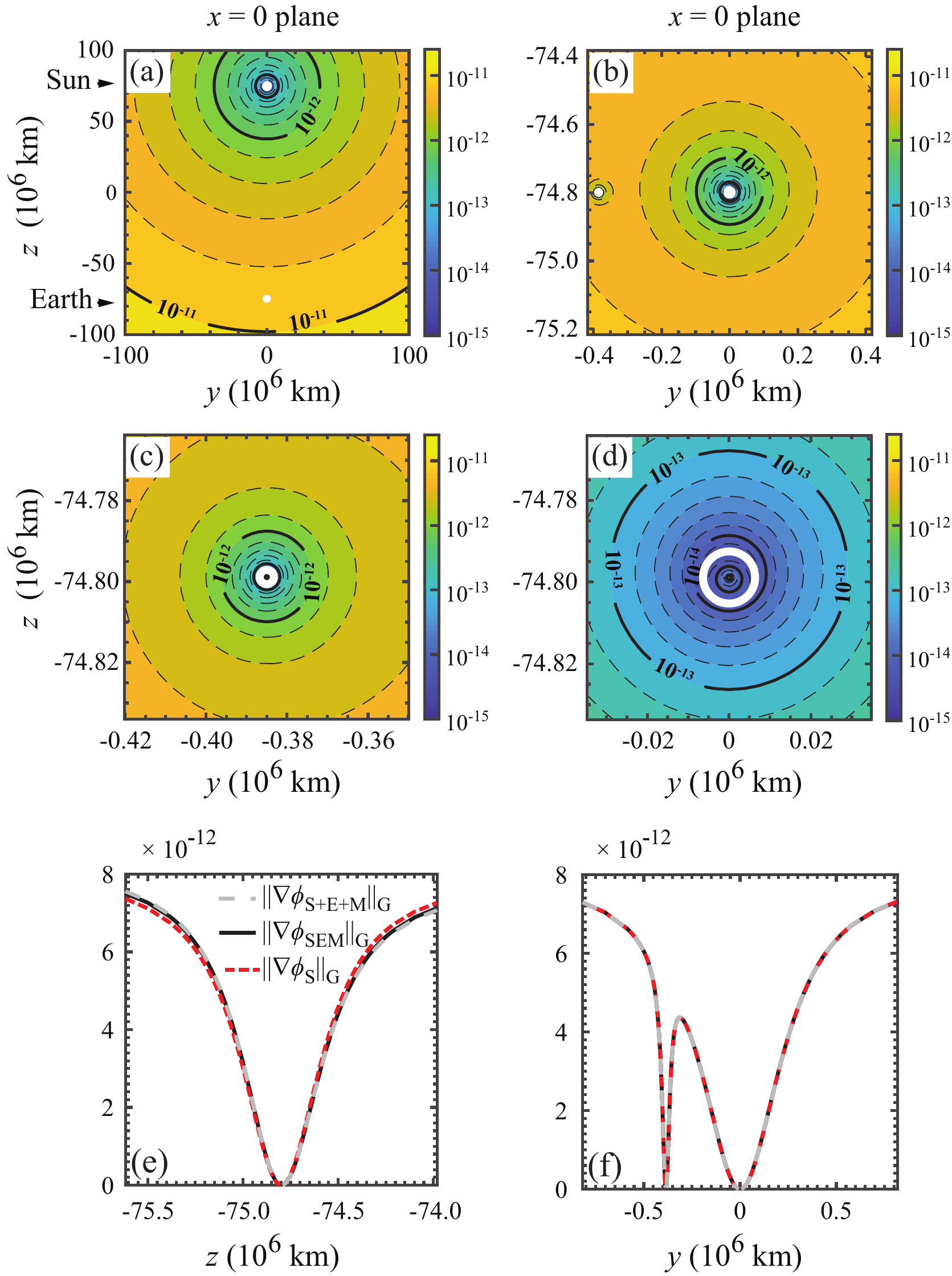}
\caption{
\label{fig:sun_earth_moon_forcestrength}
Numerical solutions for the Sun-Earth-Moon (SEM) force field  normalized by gravity (G), $\|\nabla \dimphiSEM(x,y,z)\|_G = \|\nabla \dimphiSEM(x,y,z)\|/\|\nabla \dimpsiG(x,y,z)\|$ displayed in the $x=0$ plane and on a logarithmic scale, corresponding to the simulation runs in Fig. \ref{fig:sun_earth_moon_plots_phi}. Magnitudes indicated by  solid contour lines (black) correspond to major divisions on color bar; dashed contour lines represent 1/5 intermediate color bar values. Sun, Earth and Moon bodies shown in white. The plots  exclude regions interior to the Earth and Moon bodies where the gravitational force vanishes. (a) Contour plot showing Sun (S), Earth (E) and Moon (M) bodies positioned at coordinate values ($x_{\textrm{S}}=0, y_{\textrm{S}}=0, z_{\textrm{S}}= + 74.80 \times 10^6 \textrm{km} = + 0.5 \, \textrm{AU})$, ($x_{\textrm{E}}=0, y_{\textrm{E}}=0, z_{\textrm{E}}= - 74.80 \times 10^6 \textrm{km} = - 0.5 \, \textrm{AU})$ and ($x_{\textrm{M}}=0, y_{\textrm{M}}= - 0.3845  \times 10^6 \textrm{km} = - 0.00257\, \textrm{AU}, z_{\textrm{M}}= - 74.80 \times 10^6 \textrm{km} = - 0.5 \, \textrm{AU})$. (b) Magnified view of solution in (a) centered about Earth body with the Moon to its left. (c) Magnified view of solution in (b) centered about the Moon body. (d) Magnified view of solution in (b) centered about Earth body (surface outlined in white). (e) Comparison of three solutions in the vicinity of the Earth body along the line connecting the Sun and Earth bodies: full solution $\|\nabla \dimphiSEM\|_G$ (solid black), single-body Sun solution $\|\nabla \dimphiS\|_G$ (dashed red) and combined solution $\|\dimphiSpEpM\|_G$ from linear superposition of the single-body Earth, Sun and Moon solutions (dashed gray). Span in $z$ equals a distance $8.35  \times 10^5$ km about Earth. (f) Comparison of three solutions in the vicinity of the Earth body along the line connecting the Earth and Moon: full solution, single-body Sun solution and combined solution.
}
\end{figure*}

Shown on a logarithmic scale in Fig. \ref{fig:sun_earth_moon_forcestrength} are large scale and near field views about the Earth body of the relative strength of the fifth force to the force of Newtonian gravity, $\|\nabla \dimphiASE(x,y,z)\|_G = \|\nabla \dimphiASE(r,z)\| / \|\nabla \dimpsiG\|$ for the runs shown in Fig. \ref{fig:sun_earth_moon_plots_phi}, where $\dimpsiG(r,z)$ is the Newtonian gravitational potential. The normalized values of the force in the vicinity of the Earth and Moon bodies is on the order of $10^{-12}$. As shown in (a) - (d), beyond the confines of each body, the contours are nearly spherically symmetric, with value increasing with distance from each body. The results in Fig. \ref{fig:sun_earth_moon_forcestrength} (e) and (f) depict the spatial variation in the normalized force field along the axis connecting the Sun and Earth bodies and the Moon and Earth bodies, respectively, centered about the Earth body. Curves  exclude results within the regions interior to the Earth and Moon bodies where the gravitational force vanishes. The curves indicate a very slight reduction near the Earth body and slight asymmetry about its center. Shown for comparison is the single-body Sun solution and the combined solution from linear superposition of the single-body Sun, Earth and Moon solutions. At the scales about the Earth body shown, the full numerical solution and the superposed solution are virtually indistinguishable and fairly close to the Sun solution, though the approximate solutions have opposing errors. In Fig. \ref{fig:sun_earth_moon_forcestrength} (f), all three solutions yield the same result when viewed at distances on the order of $10^6$ km.

\begin{figure*}[!h]
\centering
\includegraphics[scale=0.8]{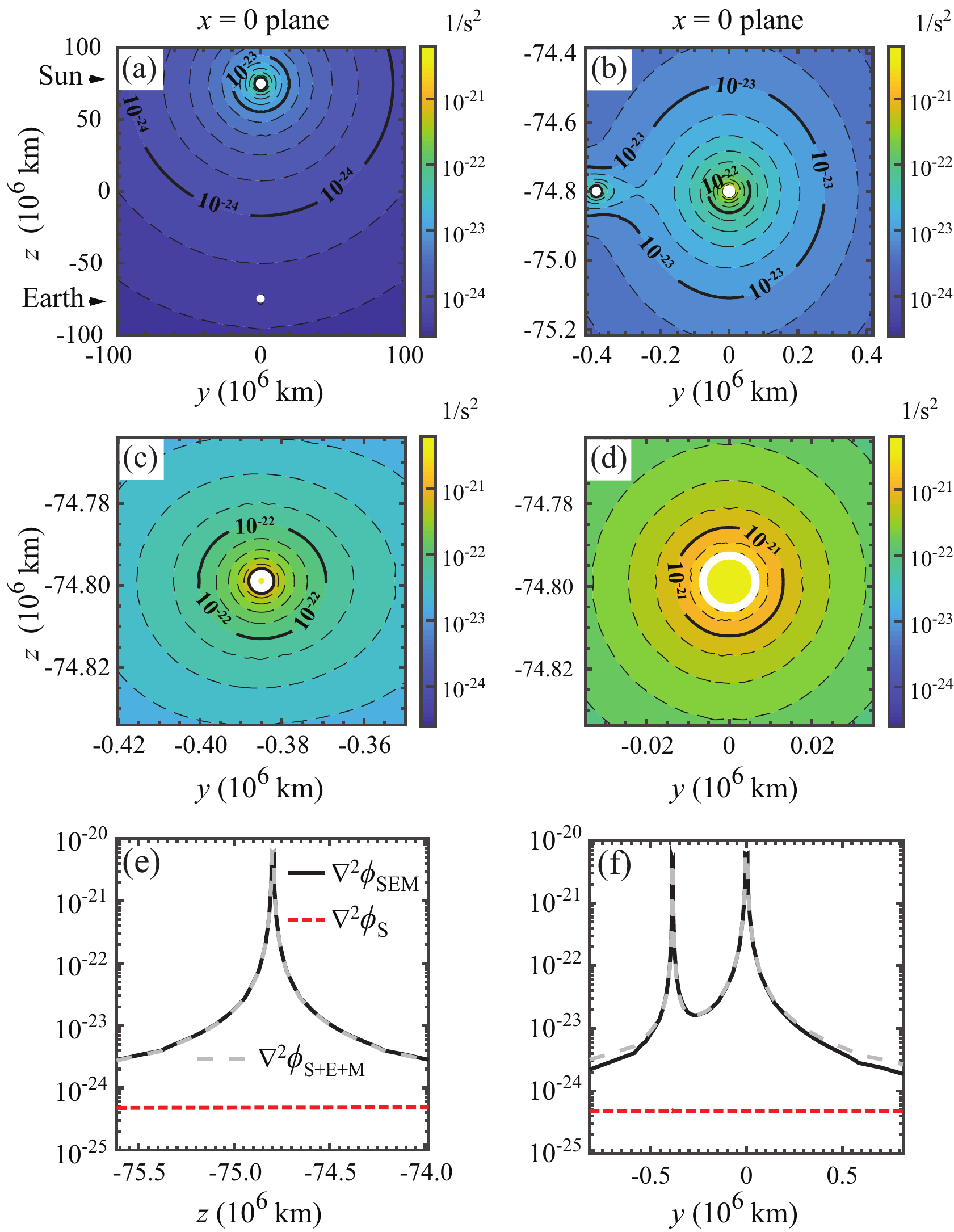}
\caption{
\label{fig:sun_earth_moon_laplacian}
Numerical solutions for the Sun-Earth-Moon (SEM) Laplacian field $\nabla^2 \dimphiSEM(x,y,z)$ $\text{s}^{-2}$ on a logarithmic scale, corresponding to the simulation runs in Fig. \ref{fig:sun_earth_moon_plots_phi}. Magnitudes indicated by solid contour lines (black) correspond to major divisions on color bar; dashed contour lines represent 1/5 intermediate color bar values. Sun, Earth and Moon bodies shown in white. (a) Contour plot in region  containing Sun (S), Earth (E) and Moon (M) bodies positioned at the coordinate values ($x_{\textrm{S}}=0, y_{\textrm{S}}=0, z_{\textrm{S}}= + 74.80 \times 10^6 \textrm{km} = + 0.5 \, \textrm{AU})$, ($x_{\textrm{E}}=0, y_{\textrm{E}}=0, z_{\textrm{E}}= - 74.80 \times 10^6 \textrm{km} = - 0.5 \, \textrm{AU})$ and ($x_{\textrm{M}}=0, y_{\textrm{M}}= - 0.3845  \times 10^6 \textrm{km} = - 0.00257\, \textrm{AU}, z_{\textrm{M}}= - 74.80 \times 10^6 \textrm{km} = - 0.5 \, \textrm{AU})$. (b) Magnified view of (a) centered about the Earth body with the Moon body to its left. (c) Magnified view of (b) centered about the Moon. (d) Magnified view of (b) centered about Earth. (e) Comparison of solutions in the vicinity of the Earth body along the line connecting the Sun and Earth: full solution $\nabla^2 \dimphiSEM$ (solid black), single-body Sun solution $\nabla^2 \dimphiS$ (dashed red) and combined solution $\nabla^2 \dimphiSpEpM$ (dashed gray) from linear superposition of the single-body Earth, Sun and Moon solutions. Span in $z$ equals a distance $8.35 \times 10^5$ km about the Earth. (f)  Comparison of three solutions in the vicinity of the Earth body along the line connecting the Moon and Earth: full solution, single-body Sun solution and combined solution.
}
\end{figure*}

Figure \ref{fig:sun_earth_moon_laplacian} shows solutions of the Sun-Earth-Moon (SEM) Laplacian field $\nabla^2 \dimphiSEM(x,y,z)$ $\text{s}^{-2}$ for the runs in Fig. \ref{fig:sun_earth_moon_plots_phi}, plotted on a logarithmic scale. The magnitudes about the Earth and Moon span roughly $10^{-23}$ to $10^{-21} \, \text{s}^{-2}$, decreasing rapidly with distance from each body. Contours of the Laplacian field along the axis connecting
the Earth and Moon in (b) and centered about the Moon in (c) exhibit some elongation. [The small ripples visible in some of the contours adjacent to the Moon and Earth surface boundaries in (c) and (d) are numerical artifacts due to meshing and not physical phenomena.] Shown also are close up views of the spatial variation in the Laplacian field in the vicinity of the Earth body along the line connecting the Sun and Earth (e) and Earth and Moon (f). For comparison, shown are the single-body Sun solution and the combined solution from linear superposition of the single-body Sun, Earth and Moon solutions. At the scales about the Earth body indicated in (e), the full numerical
solution and the solution based on linear superposition are virtually indistinguishable and differ significantly from the uniform single-body Sun solution shown. The rapid decay with increasing distance from each body in (e) accords with the single-body solution given by Eq. (\ref{eqn:spherk0}). The data in (f) indicate that the approximate solution based on linear superposition tends to overestimate the correct solution with increasing distance from the Earth and Moon bodies.

\begin{figure*}[!htb]
\centering
\includegraphics[scale=0.8]{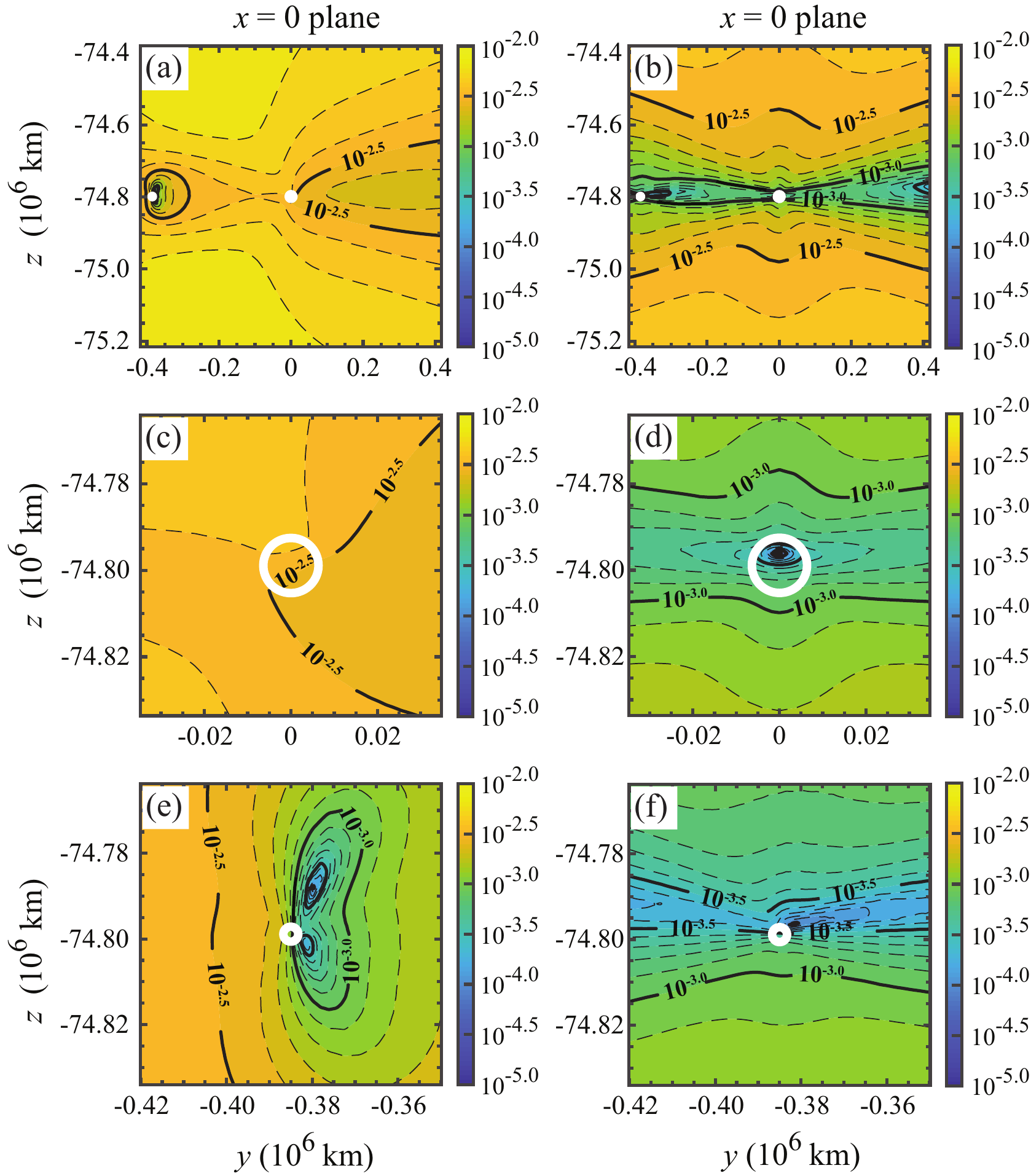}
\caption{
\label{fig:sun_earth_moon_plots_graddiff}
Contour plots showing normalized residuals of the force field - $\|\nabla \dimphiSEM - \nabla \dimphiSpEpM\|/\|\nabla \dimphiSEM\|$ (left column) and $\|\nabla \dimphiSEM -\nabla \dimphiEM -\nabla \dimphiS\|/\|\nabla \dimphiSEM\|$ (right column) - displayed in the $x=0$ plane on a logarithmic scale, corresponding to the simulation runs in Fig. \ref{fig:sun_earth_moon_plots_phi}. Magnitudes indicated by solid contour lines (black) correspond to major divisions on color bar; dashed contour lines represent 1/5 intermediate color bar values. Earth and Moon bodies shown in white. (a) and (b) Contour plots showing Earth body with Moon to its left. Body coordinates values are ($x_{\textrm{S}}=0, y_{\textrm{S}}=0, z_{\textrm{S}}= + 74.80 \times 10^6 \textrm{km} = + 0.5 \, \textrm{AU})$, ($x_{\textrm{E}}=0, y_{\textrm{E}}=0, z_{\textrm{E}}= - 74.80 \times 10^6 \textrm{km} = - 0.5 \, \textrm{AU})$ and ($x_{\textrm{M}}=0, y_{\textrm{M}}= - 0.3845  \times 10^6 \textrm{km} = - 0.00257\, \textrm{AU}, z_{\textrm{M}}= - 74.80 \times 10^6 \textrm{km} = - 0.5 \, \textrm{AU})$. (c) and (d) Magnified view of solutions in (a) and (b) centered about the Earth body (surface outlined in white). (e) and (f) Magnified view of solutions in (a) and (b) centered about the Moon body (shaded in white).
}
\end{figure*}

The contour plots depicted in Fig. \ref{fig:sun_earth_moon_plots_graddiff} show the normalized residuals for the runs in Fig. \ref{fig:sun_earth_moon_plots_phi} of the force field -- $\|\nabla \dimphiSEM - \nabla \dimphiSpEpM\|/\|\nabla \dimphiSEM\|$ (left column) and $\|\nabla \dimphiSEM -\nabla \dimphiEM-\nabla \dimphiS\|/\|\nabla \dimphiSEM\|$ (right column) -- displayed in the $x=0$ plane on a logarithmic scale. As is the case with the residual errors for the two-body Sun-Earth system shown in Fig. \ref{fig:sun_earth_plots_diffs}, the relative errors for the three-body system in Fig. \ref{fig:sun_earth_moon_plots_graddiff} are nowhere more than 1\%. These errors become even smaller when the full solution is compared against the solution based on the sum of the two-body Earth-Moon ($\phi_\textrm{EM}$) and single-body Sun solution, as evident in the right panel.

\begin{figure*}[!htb]
\centering
\includegraphics[scale=0.8]{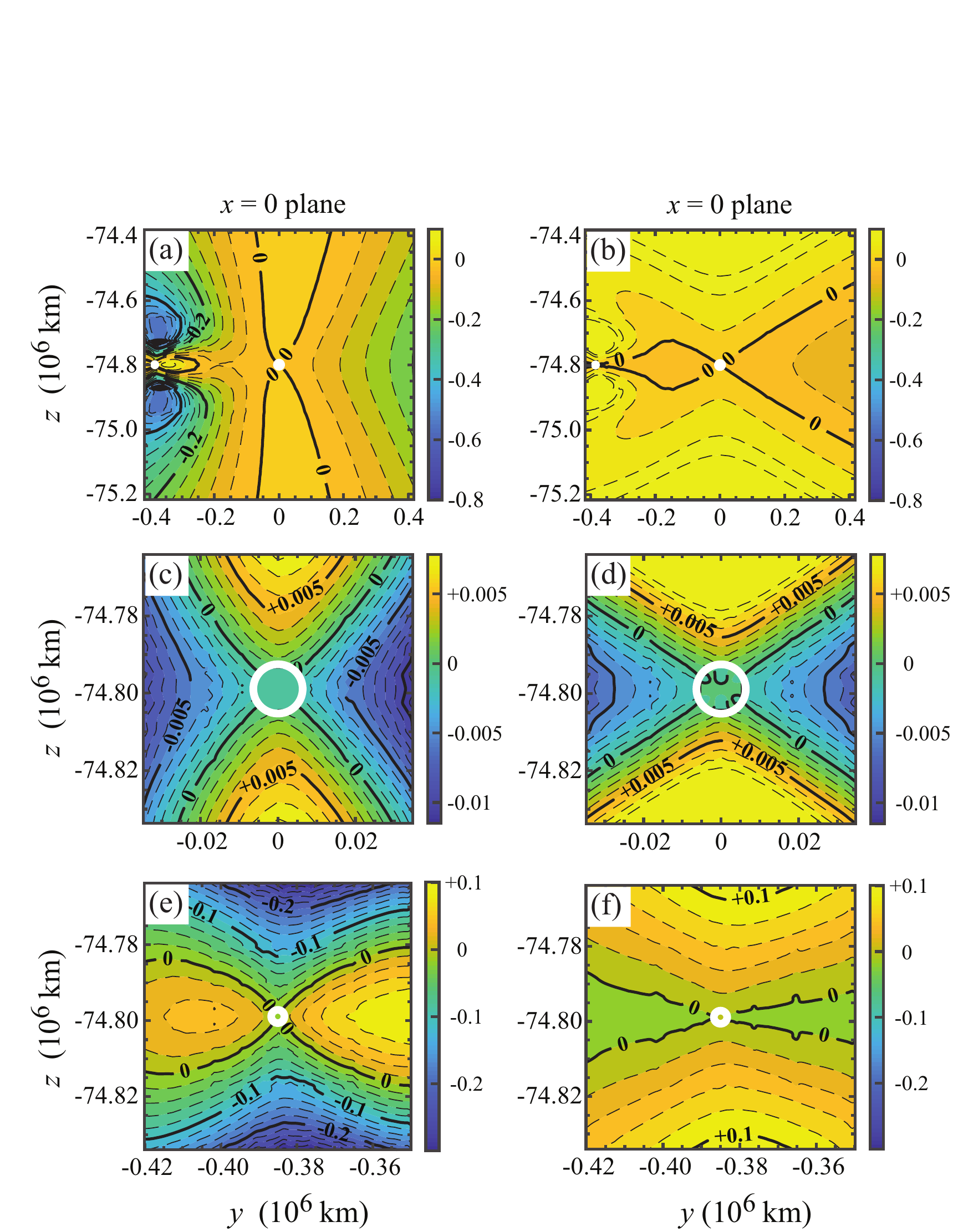}
\caption{
\label{fig:sun_earth_moon_plots_lapdiff}
Contour plots showing the normalized residuals of the Laplacian fields--$(\nabla^2 \dimphiSEM - \nabla^2 \dimphiSpEpM)/(\nabla^2 \dimphiSEM)$ (left column) and $(\nabla^2 \dimphiSEM -\nabla^2 \dimphiEM - \nabla^2 \dimphiS)/(\nabla^2 \dimphiSEM)$ (right column)--displayed in the $x=0$ plane on a linear scale, corresponding to the simulation runs in Fig. \ref{fig:sun_earth_moon_plots_phi}. Magnitudes indicated by solid contour lines (black) correspond to major divisions on color bar; dashed contour lines represent 1/5 intermediate color bar values. Earth and Moon bodies shown in white. (a) and (b) Contour plots showing Earth body with Moon to its left. Body coordinates values are ($x_{\textrm{S}}=0, y_{\textrm{S}}=0, z_{\textrm{S}}= + 74.80 \times 10^6 \textrm{km} = + 0.5 \, \textrm{AU})$, ($x_{\textrm{E}}=0, y_{\textrm{E}}=0, z_{\textrm{E}}= - 74.80 \times 10^6 \textrm{km} = - 0.5 \, \textrm{AU})$ and ($x_{\textrm{M}}=0, y_{\textrm{M}}= - 0.3845  \times 10^6 \textrm{km} = - 0.00257\, \textrm{AU}, z_{\textrm{M}}= - 74.80 \times 10^6 \textrm{km} = - 0.5 \, \textrm{AU})$. (c) and (d) Magnified view of solutions in (a) and (b) centered about the Earth body (surface outlined in white). (e) and (f) Magnified view of solutions in (a) and (b) centered about the Moon body (shaded in white).
}
\end{figure*}

The contour plots in Fig. \ref{fig:sun_earth_moon_plots_lapdiff} show the normalized residuals for the corresponding Laplacian fields - $(\nabla^2 \dimphiSEM - \nabla^2 \dimphiSpEpM)/(\nabla^2 \dimphiSEM)$ (left column) and $(\nabla^2 \dimphiSEM -\nabla^2 \dimphiEM - \nabla^2 \dimphiS)/(\nabla^2 \dimphiSEM)$ (right column). The relative errors corresponding to $(\nabla^2 \dimphiSEM - \nabla^2 \dimphiSpEpM)/(\nabla^2 \dimphiSEM)$ (left column) are of the order of 1\% within distances of about $10^5$ km from the Earth body but increase to about 80\% at distances of about $2 \times 10^5$ km from the Moon. The relative errors corresponding to $(\nabla \dimphiSEM -\nabla \dimphiEM -\nabla \dimphiS)/(\nabla \dimphiSEM)$ (right column) are smaller though still reach values of 15\% at a distances of about $2 \times 10^5$ km from the Moon. We note that the small-scale undulations visible on some of the contour lines are numerical artifacts due to the mesh size, which can be resolved by enforcing much finer meshes in the vicinity of the respective masses shown.

\clearpage
\subsection{Discussion}
The results presented were obtained from numerical simulations  of 2D axisymmetric and 3D Cartesian scalar potential fields at solar system scales in the cubic Galileon gravity model given by Eq. (\ref{eqn:vainshtein_nonlin_only}) in the limit where the coefficient of the linear Laplacian term vanishes ($k=0$). These studies for the two-body Sun-Earth system indicate that despite the nonlinearity of the governing equation, linear superposition of the individual Sun and Earth potential fields satisfying Eq. (\ref{eqn:spherk0}) for a single-body, spherically symmetric mass provides a satisfactory first-order approximation to the correct scalar field. Inspection of the corresponding differences for higher-order derivatives such as the force and Laplacian fields, both critical to experimental measurements, indicate significant deviations away from the two bodies. These results highlight that despite their sizable separation distance, the nonlinear couplings between the Earth and Sun bodies play a significant role. For the studies involving the three-body Sun-Earth-Moon system, we find that linear superposition of the individual Sun, Earth and Moon potential fields satisfying Eq. (\ref{eqn:spherk0}) for a single-body, spherically symmetric mass do not provide a satisfactory first-order approximation to the correct scalar field. Differences between the correct solution and approximations based on superposition fields for the force and Laplacian become unacceptably large.

These phenomena can be simply traced to the relatively small distance separating the Earth and Moon and the large distance separating the Earth and Sun. For the two-body Sun-Earth system, the field is dominated by the massive Sun body and its corresponding force field is practically constant in the vicinity of the Earth. This background force field (i.e., the gradient of the Sun potential field) near the Earth does not affect solutions to Eq. (\ref{eqn:nd_vainshtein_pot}) due to Galilean invariance. By contrast, the Earth and Moon are closer in mass and distance, and therefore the single-body potential field of each is stronger and the corresponding gradient functions (forces) no longer relatively constant. The two-body Earth and Moon system is therefore expected to exhibit stronger nonlinear coupling than the two-body Sun and Earth system. Indeed, the approximate solutions for the force fields and Laplacian fields based on linear superposition of the two-body Earth-Moon system and the single-body Sun solution (see right columns of Figs.  \ref{fig:sun_earth_moon_plots_graddiff} and \ref{fig:sun_earth_moon_plots_lapdiff}) show an accuracy comparable to that of the superposition solution for the two- body Sun-Earth system.

Based on our findings, we recommend that space-based detection schemes for measurements at solar system scales, which are designed around the fact that the Laplacian field for Newtonian gravity vanishes identically, will be best served by relying on predictions based on three-body Sun-Earth-Moon simulations. This will avoid potentially large errors in the range of 10-15 \% near the Earth-Moon region reported in this study. We also recommend that such detection missions be positioned in regions where the fifth force is relatively strong compared to Newtonian gravity. Our results in Fig. \ref{fig:sun_earth_moon_forcestrength} (f) indicate that this ratio achieves a local maximum between the Earth and Moon body,  corresponding to the location where their individual gravitational fields nearly cancel. This then provides an optimal location for detection of the fifth force. In fact, the results in Fig. \ref{fig:sun_earth_moon_laplacian} (b) showing significant elongation of the Laplacian field along the axis connecting the two bodies suggest that the location choice based on a local maximum in the force can also be balanced against regions exhibiting strong modulation in the Laplacian field in order to seek optimal orbits for detection and measurement.

\section{Conclusion}
\label{sec:conclusion}
In this work, we provide an accurate, stable and rapidly convergent  numerical scheme for solution of the 2D axisymmetric and 3D Cartesian scalar potential fields at solar system scales in the cubic Galileon gravity model given by Eq. (\ref{eqn:vainshtein_nonlin_only}). The method should be equally effective for non-vanishing $k$. The approach taken derives from the fact that the solar system must be treated differently from systems modeled at galactic and cosmological scales since dense mass sources have compact support and the distances relevant to solar system bodies fall well within the Vainshtein radii. 

We illustrate the numerical method by obtaining solutions for the 2D axisymmetric Sun-Earth system and 3D Cartesian Sun-Earth-Moon system. The iteration scheme is based on gradient descent of a residual function representing the positive (attractive) branch of the governing equation, which is quadratic in the Laplacian field. Due to the assumption that the dense mass sources dominate local underdensities, the algorithm  converges rapidly toward the global minimum, regardless of the initial trial solution. This behavior is confirmed by a simple analytic argument. The proposed iteration scheme is therefore robust against initial trial solutions and converges rapidly to the global minimum representing the correct two-body and three-body solutions. Generally speaking, the results of our simulations indicate that the approximate solutions based on linear superposition of fields of individual bodies may be an acceptable zero order approximation to the correct solution. But even in cases where the full 2D or 3D Galileon potential solutions do not deviate too strongly from the solutions obtained by linear superposition, higher derivatives of the scalar field, namely the force and Laplacian fields, always show unacceptable discrepancy. And since current detection schemes are being designed around measurement of the Laplacian field, we discourage use of approximate solutions based on linear superposition as a substitute for the correct solution.

Regarding the choice of boundary conditions used in such simulations, we offer the following suggestions as well. The validation studies provided in Appendix \ref{sec:numerval:bcs} offer good evidence that the far field boundary condition we applied is acceptable so long as the boundaries of the computational domain are placed sufficiently far from the location of all interior bodies. This boundary condition mimics the influence of an interior point source mass equal to the total mass of all interior bodies positioned at the center of mass of those bodies. Sensitivity studies to investigate the influence of choice of far field boundary condition should also be conducted in order to quantify how boundary perturbations affect the solution in the interior domain. In addition to this issue, even more realistic simulations can be conducted by attributing density profiles to massive bodies with spatial variation. Of course, for even more accurate predictions of the scalar, force and Laplacian fields for detection missions, even finer meshes are recommended. One could also consider a different parameter $r_c$ or include a non-unity $\beta$ term. Doing so would only multiply the results by a constant factor, giving a different estimate for the relative strength of the Galileon force and Newtonian gravity, but otherwise having no effect on our conclusions.

We anticipate that our methodology can be adopted in support of future detection missions seeking to validate the Vainshtein screening mechanism at small scales. To this end, we hope the results of this study can better guide the design of future instrumentation and bounds on precision required for such missions. To facilitate distribution of our software code and encourage further testing, we provide the following link \url{https://www.github.com/nwhite-math/small-GaPS}, where this material can be freely downloaded.

\appendix

\newcommand{\dipa}{\widehat{\partial}}
\newcommand{\diL}{\widehat{\scL}}
\newcommand{\Nmesh}{N_{\text{mesh}}}
\section{Details of implemented iteration scheme}
\label{sec:numerical_method}

\subsection{Algorithm}
\label{sec:algorithm}
We present in these two Appendices the numerical scheme used in the numerical simulations along with tests conducted to verify accuracy, stability and convergence. Note that all computations were performed in dimensionless variables according to Table \ref{table:simulationparams}; however, results are presented here in dimensional form for convenience.

The iteration scheme mentioned in Section \ref{sec:results:iteration_scheme} was carried out in MATLAB \cite{Matlab2015a} using central finite difference discretization. The mesh consisted of a discrete set of points describing a series of nested rectilinear grids described in more detail in Section \ref{sec:nested_grid_scheme}. All quantities of interest were therefore defined on mesh points. Each mesh point was specified by a unique number ranging from 1 to $\Nmesh$, the latter denoting the total number of mesh points. Each quantity of interest, such as $\ndphi$ or $\ndrho$, was stored as a vector of length $\Nmesh$, where the $i$th component defined its value at mesh point $i$.

The density field $\ndrho(\vec{r})$ for each body mass was constructed by setting all mesh points within the interior equal to the relevant density value listed in Table \ref{table:simulationparams}. All mesh points in empty space between and around bodies were set to zero. The boundary surfaces were therefore defined to within a mesh length. The initial trial solution for the non-dimensional scalar field, $\ndphi^{(n=0)}$, was then constructed from the summation of the single-body solutions obtained from Eq. (\ref{eqn:spherk0}) according to their respective masses. As discussed in Section {\ref{sec:results:iteration_scheme}, however, any other trial solution is acceptable. The values of $\ndphi^{(n=0)}$ at the boundaries were then set to the required boundary conditions. For the 2D axisymmetric Sun-Earth simulations, we applied a  Neumann condition along the symmetry axis $R=0$ such that $\partial_R \ndphi(R=0,Z) = 0$. With regard to the remaining far field boundary conditions in $R$ and $Z$, and for the far field boundary conditions chosen for the 3D Sun-Earth-Moon simulations based on Cartesian geometry, we adopted Dirichlet conditions obtained from the value of the scalar potential given by Eq. (\ref{eqn:spherk0}). Because the initial trial solution was always made to satisfy the boundary condition, iterative corrections were computed using homogeneous boundary conditions.

Discrete differential operators $\dipa_r$, $\dipa_r^2$, $\dipa_z$, and $\dipa_z^2$ for the 2D axisymmetric simulations and $\dipa_x$, $\dipa_x^2$, $\dipa_y$, $\dipa_y^2$, $\dipa_z$, and $\dipa_z^2$ for the 3D Cartesian simulations (where $\dipa$ denotes the discrete version of $\partial$) were constructed according to the central difference scheme described in Section \ref{sec:nested_grid_scheme}. Each of these operators was stored as an $\Nmesh \times \Nmesh$ matrix.

At each iteration step $n$, the corresponding discrete residual function for Eq. (\ref{eqn:residual}) given by
\begin{align}
  \notag
  \scR^{(n)} &=
  \sqrt{\sum_{i,j} \left(\dipa_{i}\dipa_{j} \ndphi^{(n)} \right)^2 + \ndrho + \left( \frac{k}{2} \right)^{2}}~-
  \\
  \label{eqn:discresidual}
  &\phantom{=}~
  \sum_i \dipa_i^2 \ndphi^{(n)} - \frac{k}{2}
  ,
\end{align}
and the discrete linear operator for Eq. (\ref{eqn:linear_operator}) given by
\begin{align}
  \diL^{(n)} &=
  \frac{\sum_{i,j} \left( \dipa_i \dipa_j \ndphi^{(n)} \right) \dipa_i \dipa_j}{\sqrt{\sum_{\ell,m} \left(\dipa_{\ell}\dipa_{m} \ndphi^{(n)} \right)^2 + \ndrho + \left( \frac{k}{2} \right)^{2}}} - \sum_i \dipa_i^2
\end{align}
were computed. Here we include the linear term with coefficient $k$. For $i \ne j$, the term $\dipa_i \dipa_j$ was computed by matrix multiplication since even in discrete form, the product is commutative. The term $\dipa_i^2$ was computed using its own stencil instead of multiplying together the two first-order derivative operators. We refer to Section \ref{sec:nested_grid_scheme} for further explanation.
Like the $\dipa_i$ operator matrix, the linear operator $\diL^{(n)}$ was also stored as an $\Nmesh \times \Nmesh$ matrix.

The correction step $\xi^{(n)}$ was then computed by solving the equation
\begin{align}
  \label{eqn:schematic_LxiR}
  \diL^{(n)} \xi^{(n)} &= - \scR^{(n)}
  .
\end{align}
This step, based on a linear solver, is described in more detail below. The correction $\xi^{(n)}$ was then added to the current value of $\ndphi^{(n)}$ to yield the updated solution $\ndphi^{(n+1)}$, namely
\begin{align}
  \ndphi^{(n+1)} = \ndphi^{(n)} + \nu \xi^{(n)}
  .
\end{align}
Had the classical Newton-Raphson method been used instead, the gradient step size $\nu$ would have equalled 1 \cite{H53}, but convergence would not have been guaranteed. Dynamically reducing $\nu$ to be less than 1 ensured that the integrated residual decreased at every iteration \cite{B65}. In the present implementation, $\nu$ was chosen to be 1 whenever possible. If $\int (\scR^{(n+1)}[\ndphi^{(n)} + \xi^{(n)}])^2 dV > \int (\scR^{(n)}[\ndphi^{(n)}])^2 dV$ (i.e., the residual error did not decrease), then $\nu$ was halved to a value of 0.5. If this smaller step size still did not reduce the residual error, $\nu$ was halved yet again. In this manner, the step size $\nu$ was continually decreased by powers of two until either the residual decreased or attained a limiting value of $10^{-10}$. Once a value of $\nu$ was found which successfully reduced the residual, the iteration loop was allowed to continue, i.e., $\scR^{(n+1)}$, $\diL^{(n+1)}$, $\xi^{(n+1)}$, etc were constructed. If no step size $\nu$ could be found which reduced the residual, then the iteration loop was either aborted or switched to a different linear solver, as described below.

The boundary conditions were handled in two different ways. Whenever the iteration loop for a minimum step size $\nu$ did not reduce the value of the residual, the algorithm was switched to an alternate linear solver that applied the boundary conditions differently. This approach was found to improve the final value of the residual by a few percent in comparison to results obtained using either solver alone.

Some additional notation is required before describing these linear solvers. Let $B \subset \{1, \dots,\Nmesh\}$ denote the set of mesh points on the boundary of the computational domain, and $I$ denote the mesh points within the domain interior, so that $I \cup B = \{1, \dots, \Nmesh\}$. Let square brackets denote indexing, so that for example, $\xi^{(n)}[I]$ denotes the subvector of $\xi^{(n)}$ defined on interior mesh points and $\diL^{(n)}[I \cup B,I]$ denotes the rectangular submatrix of $\diL^{(n)}$ consisting of rows corresponding to all nodes and columns corresponding only to interior nodes.

The first linear solver relied only on the interior points such that
\begin{align}
\notag
  \xi^{(n)}[B] &= 0
  ,
  \\
  \diL^{(n)}[I,I] \xi^{(n)}[I] &= -\scR^{(n)}[I]
  .
\end{align}
Since the matrix $\diL^{(n)}[I,I]$ is square and invertible, a solution was guaranteed, which was obtained using the direct solver in MATLAB \texttt{mldivide} based on least squares. This was the approach taken for most of the runs conducted. For cases involving large 3D meshes, the iterative biconjugate gradient solver \texttt{bicgstab} in MATLAB was used instead, with the diagonal of $\diL^{(n)}[I,I]$ used as a preconditioner. When the process \texttt{bicgstab} failed to converge, the algorithm was made to revert back to the direct solver \texttt{mldivide}. The second linear solver relied on the fact that $\scR^{(n)}$ is defined on both interior and boundary nodes such that the equation could be solved immediately as a least squares problem using \texttt{mldivide}, according to which
\begin{align}
\notag
  \xi^{(n)}[B] &= 0
  ,
  \\
  \xi^{(n)}[I] &= \argmin_{\xi^{(n)}[I]} \left(\diL^{(n)}[I \cup B,I] \xi^{(n)}[I] + \scR^{(n)}[I \cup B]\right)^2
  .
\end{align}

\subsection{Nested grid finite difference scheme}
\label{sec:nested_grid_scheme}
One of the challenges in simulating the scalar potential field over solar system distances is the range of length scales which must be resolved numerically. For example, the radius of the Sun is approximately $5 \times 10^{-3}$ AU, while the radius of Earth is only about $4 \times 10^{-5}$ AU. Constructing a uniform 3D rectilinear mesh covering one cubic AU, with mesh spacing of one Earth radius, would easily demand about $10^{13}$ points, clearly not an effective use of computational resources. One alternative is to construct a rectilinear mesh with variable mesh spacing, the approach used by Hiramatsu \textit{et al.} \cite{HS13}. Constraining variable mesh spacings to be rectilinear, however,  inevitably leads to distorted spacings of high aspect ratio in regions where the mesh is fine along one coordinate axis but coarse along another. When possible, it is preferable instead to implement local mesh refinement.

To resolve this issue without introducing an entirely unstructured mesh, a system of nested rectilinear meshes was employed. This choice led to two types of mesh points: interior points which were not on a boundary between coarse and fine regions, and boundary points. Derivatives on interior points were then computed at second order using a 3-point central difference scheme, while boundary points involved a more complex stencil to include interpolated ``halo points'' \cite{FL19}. A diagram outlining this nested mesh scheme (confined to 2D for simplicity) is shown in Fig. \ref{fig:multimesh_stencil}. The solid circles (blue) denote mesh points on a fine mesh with spacing of $h$, the solid squares (red) denote mesh points on an exterior coarser mesh with spacing of $2h$, and the open diamonds (white) denote interpolated halo points.

We illustrate this scheme for the 3D Cartesian system. Let $x$, $y$ and $z$ coordinates be indexed by $i,j,k$ so that $\{i+1,j,k\}$ is the point immediately adjacent to $\{i,j,k\}$ along the $x$-axis. Let $f_{i,j,k}$ denote the value of a scalar function $f$ on the mesh point $\{i,j,k\}$ where $x_{i,j,k}$ denotes the value of the coordinate $x$ at that point, and so on. Let then $x_{i,j,k} - x_{i-1,j,k} = h_1$ and $x_{i+1,j,k} - x_{i,j,k} = h_2$. In Fig. \ref{fig:multimesh_stencil}, $h_1=h$ and $h_2=2h$.

All derivatives were computed to second order. Variations in the scale function $f$ along the $x$ axis, for example, are given by \cite{SV70}:
\begin{align}
  \frac{\dipa f}{\dipa x} &= \frac{h_1^2\left(f_{i+1,j,k} - f_{i,j,k}\right) + h_2^2\left(f_{i,j,k} - f_{i-1,j,k}\right)}{h_1 h_2 (h_1 + h_2)}
  ,
  \\
  \frac{\dipa^2 f}{\dipa x^2} &= 2\frac{h_1 \left( f_{i+1,j,k} - f_{i,j,k}\right) - h_2 \left(f_{i,j,k} - f_{i-1,j,k}\right)}{h_1 h_2(h_1 + h_2)}
  ,
\end{align}
and similarly for $y$ and $z$. On the interior of each submesh, points are equispaced and the derivatives reduce to central difference. At the outer edges of the outermost mesh, derivatives are computed to $O(h^2)$ using two neighboring points. For example, letting $x_{2,j,k} - x_{1,j,k} = h_2$ and $x_{3,j,k} - x_{1,j,k}= h_3$,
\begin{align}
  \frac{\dipa f}{\dipa x} &= \frac{1}{h_3 - h_2}\left[ h_3 \frac{f_{2,j,k} - f_{1,j,k}}{h_2} - h_2 \frac{f_{3,j,k} - f_{1,j,k}}{h_3} \right]
  ,
  \\
  \frac{\dipa^2 f}{\dipa x^2} &= \frac{2}{h_3 - h_2}\left[ \frac{f_{3,j,k} - f_{1,j,k}}{h_3} - \frac{f_{2,j,k} - f_{1,j,k}}{h_2} \right]
  .
\end{align}
Similarly for derivatives along the $y$ and $z$ axes.

The above scheme, of course, relies on every point having neighboring points. However, certain points on the boundary of fine submeshes will not have a neighboring point in the exterior mesh. In Fig. \ref{fig:multimesh_stencil} for example, the mesh point $\{i,j,k\}$ has no neighbor to the right along the $x$ axis and $\{i-1,j+1,k\}$ has no neighboring point above it along the $y$ axis. To compute a second order $x$ derivative at $\{i,j,k\}$, for example, information from the surrounding points $\{i,j+1,k\}$, $\{i+1, j+1, k\}$, $\{i,j-1,k\}$, $\{i+1,j-1,k\}$ and $\{i-1,j,k\}$ is required. The information from all these surrounding points can be incorporated through the introduction of a halo point at $\{i+1, j, k\}$. To illustrate this from Fig. \ref{fig:multimesh_stencil}, the halo point is defined as
\begin{align}
  \notag
f_{i+1,j,k} &= f_{i,j,k} + 2 h \times
\\& \frac{1}{2}\left[ \frac{f_{i+1,j+1,k}-f_{i,j+1,k}}{2h} + \frac{f_{i+1,j-1,k}-f_{i,j-1,k}}{2h} \right]
.
\end{align}
The halo point is therefore defined by linear interpolation of nearby points. In particular, the $x$-derivative is approximated by a weighted average of first order derivatives at $f_{i,j-1,k}$ and $f_{i,j+1,k}$, and the result multiplied by $2 h$ to extrapolate from $f_{i,j,k}$ to $f_{i+1,j,k}$. The 3D analogue is identically computed except that the neighboring points $f_{i,j,k-1}$ and $f_{i,j,k+1}$ along the $z$ axis are also used in the weighted average. Multiple derivatives for different variables, such as $\dipa^2 f/(\dipa x \dipa y)$, are constructed directly by first computing $\dipa f / \dipa x$ and then $\dipa/\dipa y$ at each mesh point.

\begin{figure}[!htb]
\centering
\includegraphics{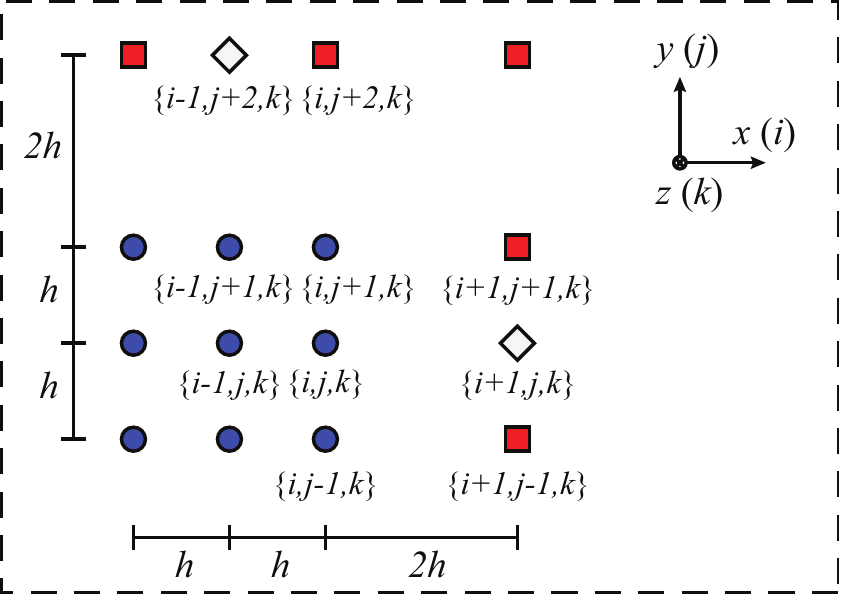}
\caption{
\label{fig:multimesh_stencil}
Diagram of nested mesh structure used in numerical simulations. Filled circles (blue) indicate points on a fine mesh with periodic spacing $h$. Filled squares (red) indicate points on an exterior coarser mesh with periodic spacing $2h$. Open diamonds (white) indicate interpolated halo points.
}
\end{figure}

\subsection{Meshes used in simulations}
\label{sec:numerical_method:meshes}
The 2D axisymmetric Sun-Earth simulations included 26 nested rectangular meshes, each twice as long in the $\hat{z}$ direction compared to the $\hat{r}$ direction. Each mesh consisted of $(n+1)$ points per side along $\hat{r}$ and $(2n+1)$ points per side along $\hat{z}$, with $n=32$ for all but one of the submeshes which contained $n=64$. The submeshes were divided into 7 outer meshes centered at the midpoint of the Sun and Earth bodies and containing both, 6 Sun body centered meshes containing only the Sun, and 13 Earth body centered meshes containing only Earth. The outer meshes extended over a radial distance from the origin equal to $2^p$ AU and a total longitudinal distance equal to $2^{p+1}$ AU for $0 \leq p \leq 6$. The system for $p=0$ required extra mesh points ($n=64$) since there was no outer mesh of size 0.5 AU, as such a mesh would have bifurcated the Sun and Earth bodies. The Sun-centered meshes were constructed to have a radial range of $2^{-2}$ AU, $2^{-3}$ AU, \ldots, $2^{-7}$ AU, the last representing a distance roughly 1.7 times the radius of the Sun. The Earth-centered meshes were constructed to have a radial range of $2^{-2}$ AU, \ldots, $2^{-14}$ AU, the last roughly 1.4 times the radius of Earth.

The 3D Cartesian Sun-Earth-Moon simulations were constructed  similarly and included 32 nested cubic meshes, each with $(n+1)$ points per side along each of the $\hat{x}$, $\hat{y}$ and $\hat{z}$ axes. All but two of the meshes were designed with $n=10$; two submeshes were designed with $n=20$. The submeshes consisted of the following collection: 7 outer meshes centered at the midpoint of the Sun and Earth bodies, 6 centered about the Sun containing only the Sun, 7 centered about the Earth containing both the Earth and Moon bodies, 5 additional meshes centered about the Earth containing only the Earth body, and 7 centered about the Moon containing only the Moon. The outer meshes had side lengths $2^{p+1}$ AU for $0 \leq p \leq 6$. The $p=0$ system required extra mesh points ($n=20$), again due to the fact that there was no outer mesh of side length 1 AU.
The Sun-centered meshes had side lengths $2^{-1}$ AU, $2^{-2}$ AU, \ldots, $2^{-6}$ AU, the last roughly 1.7 times the diameter of the Sun. The Earth-centered meshes had side extent $2^{-1}$ AU, \ldots, $2^{-7}$ AU and $2^{-9}$ AU, \ldots $2^{-13}$ AU, the last roughly 1.4 times the diameter of the Earth. The choice $2^{-8}$ AU was not implemented, since the edge of such a mesh would have bifurcated the Moon body. Instead, double the number of mesh points was used for the runs with side lengths $2^{-7}$ AU. The Moon-centered meshes had side lengths $2^{-9}$ AU, \ldots, $2^{-15}$ AU, the last roughly 1.3 times the diameter of the Moon.

\section{Validation and benchmarking of numerical algorithm}
\label{sec:numerval}
The analysis in Section \ref{sec:results:iteration_scheme} describes the iteration scheme from an analytic standpoint, and the proofs therein cannot be applied exactly to a discretized approximation. That said, we observed fast convergence even in the finite difference implementation and encountered no numerical instabilities. In this section, we provide results of numerical tests to validate the implementation of our algorithm.

\subsection{Solution convergence study}
\label{sec:numerval:fastconvergence}
The arguments presented in Section \ref{sec:results:iteration_scheme} indicate that the numerical simulations should converge rapidly regardless of choice of initial trial function for the scalar field potential. Convergence tests were therefore conducted to quantify approach to the global minimum representing the solution to Eq. (\ref{eqn:vainshtein_nonlin_only}). A variety of initial trial solutions was tested which included a uniform zero field, as well as nine distributions representing both white and red noise, each initiated from a different seed.

The (non-dimensional) white noise trial function was represented by values on each mesh node extracted from a normal distribution with zero mean and a standard deviation of 270.3, reflecting the range in values of the single body Sun solution $\ndphi(R)$ given by Eq. (\ref{eqn:spherk0}) evaluated within a distance of 512 AU from the Sun body. The (non-dimensional) red noise trial function was represented by
\begin{align}
  \ndphi^{(n=0)} = 270.3\sum_{j=1}^{100} a_j \prod_{X_i = \{X,Y,Z\}} \frac{\sin(\kappa_{j,i} X_i + \theta_{j,i})}{\sqrt{\kappa_{j,X}^2 +\kappa_{j,Y}^2+\kappa_{j,Z}^2}}
.
\end{align}
Here, $j$ denotes the 100 wave numbers along each coordinate direction selected uniformly from a logarithmic distribution ranging from $10^{-3} - 10^{3}$ where the wave numbers for the 2D axisymmetric case are labeled $\kappa_{j,R}$ and $\kappa_{j,Z}$ and for the 3D Cartesian case $\kappa_{j,X}$, $\kappa_{j,Y}$ and $\kappa_{j,Z}$. The corresponding amplitudes $a_j$ were chosen from a normal distribution with zero mean and normalized to unity such that $\sum_j a_j^2 = 1$. The phase offsets represented by $\theta_{j,X}$ and the like were chosen uniformly from the range $[0, 2\pi]$.

Figure \ref{fig:convergence} shows results of the volume averaged integration of the dimensionless residual error squared computed after each iteration step $n$ according to Eq. (  \ref{eqn:discresidual}). Within just a few iterations, the integrated residual decays rapidly by many orders of magnitude, followed by a second substantial drop, and is observed to asymptote rapidly to values below $10^{-4}$. Indeed, the results of Figure \ref{fig:convergence} confirm that in both the 2D and 3D systems studied, the integrated residual error for all test cases converged to the same small value within no more than 25 iterations.

\begin{figure}[!htb]
\centering
\includegraphics[scale=0.9]{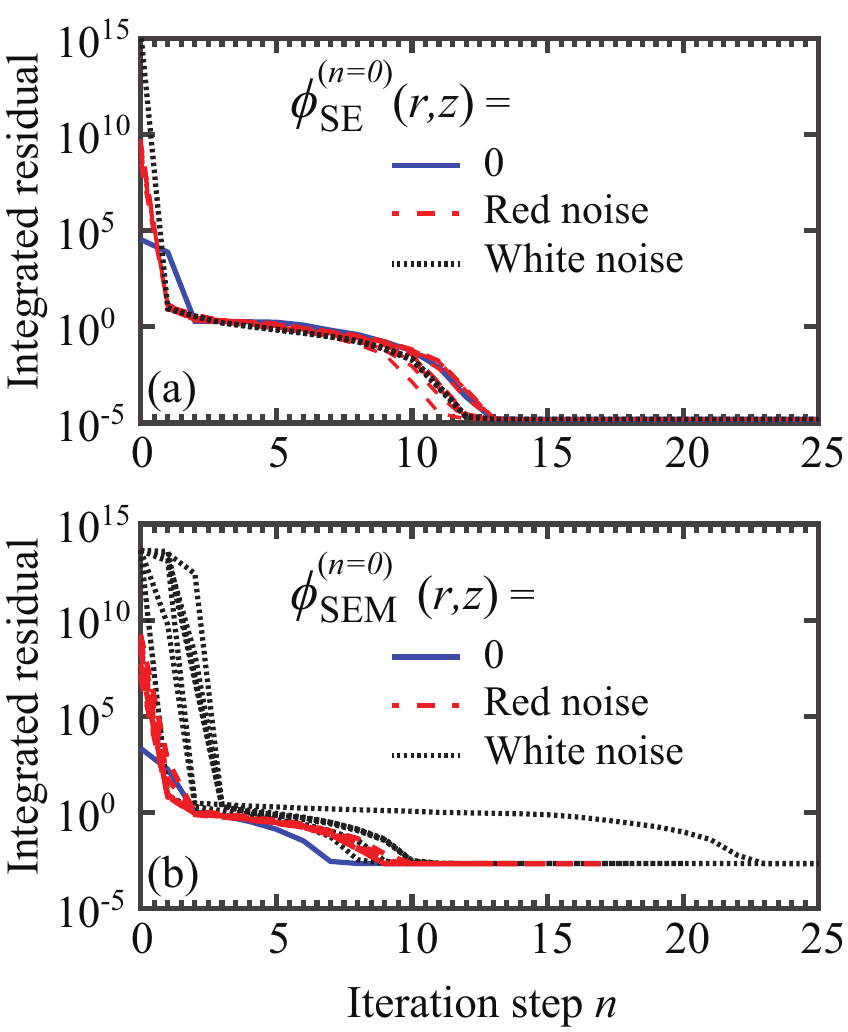}
\caption{
\label{fig:convergence}
Results of volume averaged integration of the (dimensionless) residual error squared $d^{-3}\int(\scR^{(n)}[\ndphi])^2 dV$ computed after each iteration step $n$ according to Eq. (\ref{eqn:discresidual}). (a) Results for 2D  axisymmetric Sun-Earth potential field $\dimphiASE(r,z)$. (b) Results for Sun-Earth-Moon potential field $\dimphiSEM(x,y,z)$. Shown are three types of initial trial functions: $\phi^{(n=0)}=0$ (solid blue line), red noise (long dashed red line) and white noise (short dashed black line). White noise and red noise distributions were generated from nine different seeds each. Further detail provided in Section \ref{sec:numerval:fastconvergence}.
}
\end{figure}

Although the analytical argument suggests that the integrated residual error should rapidly decay to zero, this cannot occur, or course, since the solution domain is represented by a discretized mesh. Because all points on domain boundaries are fixed by Dirichlet boundary conditions and only internal mesh points are free to vary, there are not enough degrees of freedom to achieve a pointwise residual of zero. Furthermore, the gradient descent step computed from the equation $\scL[\ndphi] \xi = -\scR[\ndphi]$ is a discrete approximation. However, higher accuracy can be achieved by implementation of other higher order finite difference schemes on even finer meshes than the basic implementation outlined in Appendix \ref{sec:nested_grid_scheme}.

\subsection{Finite size study}
\label{sec:numerval:bcs}
Far beyond the Vainshtein radius of the largest body in a collection of bodies, the Galileon scalar potential is expected to vanish \cite{CT13} such that $\lim (R \to \infty)\ndphi(\vec{R}) = 0$.
In contrast to previous studies \cite{HS13}, our computational domain falls well within the Vainshtein radii of all included bodies, and we therefore argue that it is natural to apply the approximate boundary condition set by the values of the Galileon field given by Eq. (\ref{eqn:spherk0}). This seems a valid choice so long as all computational boundaries are positioned at distances far greater than any internal length scales such as body separation distances. To validate this choice and to quantify finite size effects, we carried out simulations with domain boundaries positioned increasingly distant from the massive bodies. These simulations were carried out for the 2D axisymmetric Sun-Earth and 3D Cartesian Sun-Earth-Moon systems, which are the subject of the current work, as well as the idealized two-body system investigated by Hiramatsu \textit{et al.} \cite{HS13}. The origin of each coordinate system was positioned halfway between the two bodies for the idealized cases, and halfway between the Sun and Earth bodies for the solar system cases.
\begin{figure}[!h]
\centering
\includegraphics{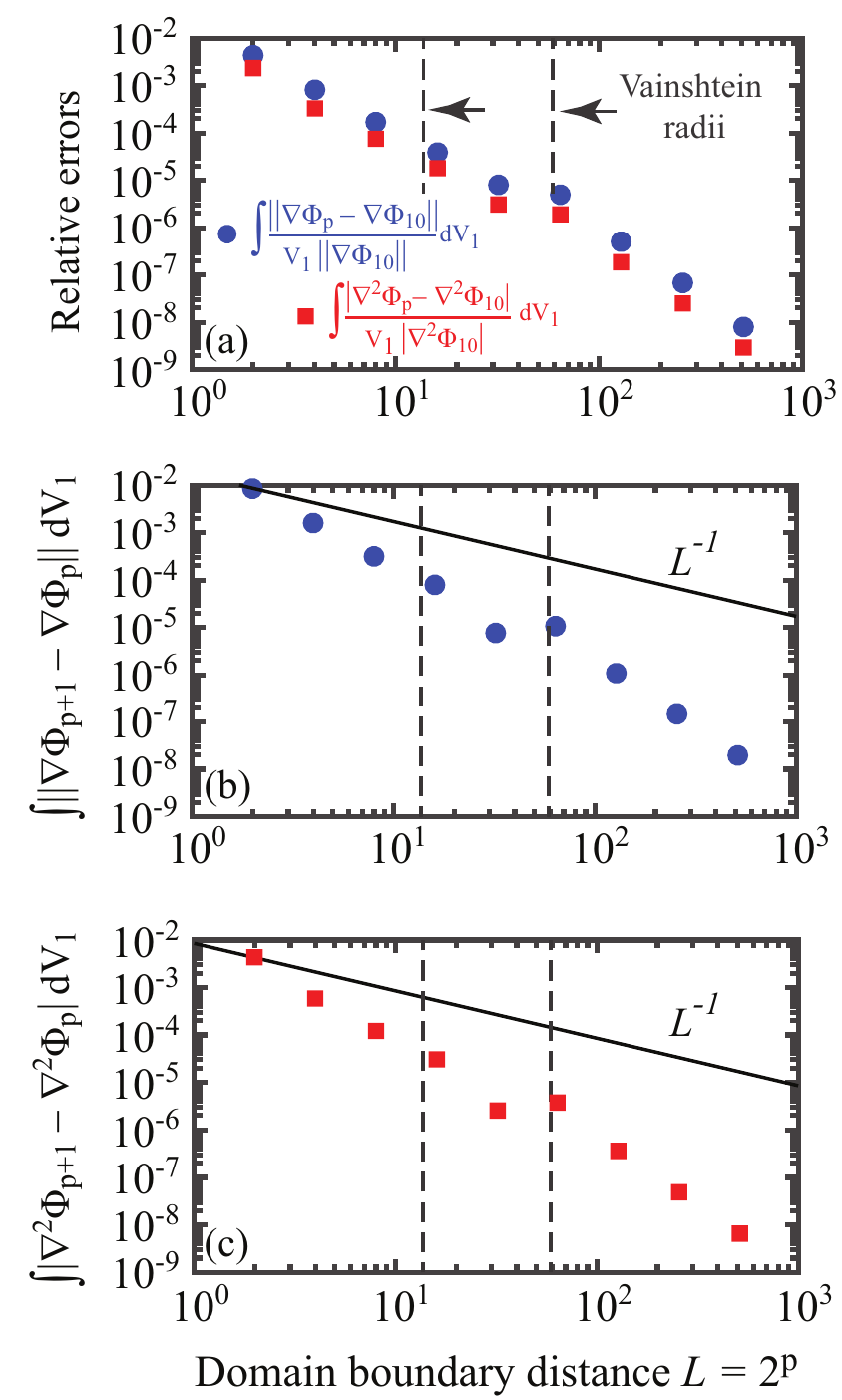}
\caption{
\label{fig:distant_boundary_convergence_hira}
Results of convergence tests carried out in a cylindrical domain for increasing domain boundary distance $L$ for the Galileon field of an idealized axisymmetric two body system. The parameter values were obtained from the study by Hiramatsu \textit{et al.} \cite{HS13} according to which the dimensionless radii and densities of the two bodies equalled $(0.3, 1.0)$ and $(0.1, 0.3375)$, the separation distance equalled one, $k=5.93 \times 10^{-4}$ and the Vainshtein radii (vertical dashed lines) equalled $58.9$ and $13.7$, respectively. Additional details can be found in Section \ref{sec:numerval:bcs}. (a) Log-log plot showing the volume averaged relative errors in field strength $\int\|\nabla \ndphi_{\textrm{p}} - \nabla \ndphi_{10}\|/(V_1 \|\nabla \ndphi_{10}\|) dV_1$ (solid blue circles) and Laplacian field $\int |\nabla^2 \ndphi_{\textrm{p}} - \nabla^2 \ndphi_{10}|/(V_1 |\nabla^2 \ndphi_{10}|) dV_1$ (solid red squares) for increasing domain size $L^{2\textrm{p}}$ for integer values $1 \leq \textrm{p} \leq 10$. (b) Log-log plot of $\int\|\nabla \ndphi_{\textrm{p}+1} - \nabla \ndphi_{\textrm{p}}\| dV_1$ for increasing domain size. (c) Log-log plot of $\int |\nabla^2 \ndphi_{\textrm{p}+1} - \nabla^2 \ndphi_{\textrm{p}}| dV_1$ for increasing domain size. Shown in (b) and (c) for comparison is the decay function $L^{-1}$.
}
\end{figure}

In Fig. \ref{fig:distant_boundary_convergence_hira}, our results for the same two-body system examined by Hiramatsu \textit{et al.} are plotted in non-dimensional form. In Fig. \ref{fig:distant_boundary_convergence} for the Sun-Earth (SE) and Sun-Earth-Moon (SEM) systems, our results are plotted in  dimensional form for the convenience of experimentalists. The  non-dimensional length scale $L=2^\textrm{p}$ refers to the radius of a cylindrical domain of volume $V_\textrm{p} = \pi \times (2^\textrm{p})^2 \times 2^{\textrm{p}+1}$ for integer values $1 \leq \textrm{p} \leq 10$ used to compute the non-dimensional Galileon field potential $\ndphi_p(R,Z)$. The dimensional length scale  $\ell=2^\textrm{p} \, \textrm{AU}$ refers instead to the distance from the origin of the computational domain to its nearest boundary. For those simulations carried out in cylindrical domains, this distance $\ell$ equaled the radius of a cylinder of volume $V_\textrm{p} = \pi \times (2^\textrm{p})^2 \times 2^{\textrm{p}+1} \, \textrm{AU}^3$. For simulations carried out in a cubic domain, this distance $\ell$ equaled the half length of the edge of a cube of volume $V_\textrm{p} = 2^{\textrm{p}+1} \times 2^{\textrm{p}+1} \times2^{\textrm{p}+1} \, \textrm{AU}^3$ for integer values $1 \leq \textrm{p} \leq 9$. The actual simulations to determine the potential fields $\phi_\textrm{p}(\vec{r})$ m$^2$/s$^2$ for the SE and SEM systems were, of course, carried out in dimensionless coordinates, with the results then plotted in dimensional form.  For proper comparison, all differences reported were evaluated only within the smallest volume common to all volumes tested for a given system, namely $V_{\textrm{p}=1}$. All relative errors are reported in comparison to the solutions obtained for the largest domain size tested.

For the idealized two-body system, we used the parameter values given by Hiramatsu \textit{et al.} \cite{HS13}. Accordingly, body A was assigned a radius and density of $(0.3, 1.0)$, respectively, and body B was assigned the values $(0.1, 0.3375)$. The two bodies were given a separation distance of $1.0$. These choices yielded a non-dimensional value for the linear coefficient in Eq. (\ref{eqn:nd_vainshtein_pot}) $k = 5.93 \times 10^{-4}$ and Vainshtein radii of $58.9$ and $13.7$, respectively. The results in Fig. \ref{fig:distant_boundary_convergence_hira}(a) demonstrate just how small are the relative errors for the force and Laplacian fields when compared to the results for the largest domain. The comparison in (b) and (c) of the results for the gradient and Laplacian fields to the decay function $L^{-1}$ also confirm rapid convergence. The results in (b) showing the mean relative difference in the force field for the smallest domain $V_1$ (where $L$ falls well within the Vainshtein radii) and the largest domain $V_{10}$ (where $L$ far exceeds the  Vainshtein radii) is only about 0.43\%. The corresponding mean relative difference for the Laplacian field, shown in (c), is only about  0.23 \%. There does appear a region around the larger Vainshtein radius at which the convergence stalls, but the relative error subsequently continues to decrease as the size of the computational domain increases. A more comprehensive study of boundary conditions is required to determine whether this stall is spurious.

The results in Fig. \ref{fig:distant_boundary_convergence} show convergence of the solar system simulations with increasing domain size. The Sun-Earth (3D Cartesian) and Sun-Earth-Moon (3D Cartesian) results are indistinguishable to two significant digits. The results in (a) and (b) demonstrate rapid convergence with increasing $\ell$ when compared to the decay function $\ell^{-1}$. The results in (c) and (d) evidence numerical consistency with increasing $\ell$, as expected. Quantitatively, in the SE (cylindrical) simulations, the mean relative gradient difference between the simulations carried out with $\ell = 2 \, \textrm{AU}$ and $\ell = 256 \,  \textrm{AU}$ was only 0.075\% and between the $\ell = 64 \, \textrm{AU}$ and $\ell = 256 \,  \textrm{AU}$ simulations only 0.00033\%. The corresponding mean relative Laplacian difference was 0.16\% and 0.000060\%, respectively. Likewise for the  Sun-Earth (3D Cartesian) and Sun-Earth-Moon (3D Cartesian) simulations, the mean relative gradient difference between the simulations carried out with $\ell = 2 \, \textrm{AU}$ and $\ell = 256 \,  \textrm{AU}$ was only $0.33$\% and between the $\ell = 64 \, \textrm{AU}$ and $\ell = 256 \,  \textrm{AU}$ simulations only $0.00071$\%. The corresponding mean relative Laplacian difference was 1.7\% and 0.0010\%, respectively.

\begin{figure*}[!h]
\centering
\includegraphics[scale=0.8]{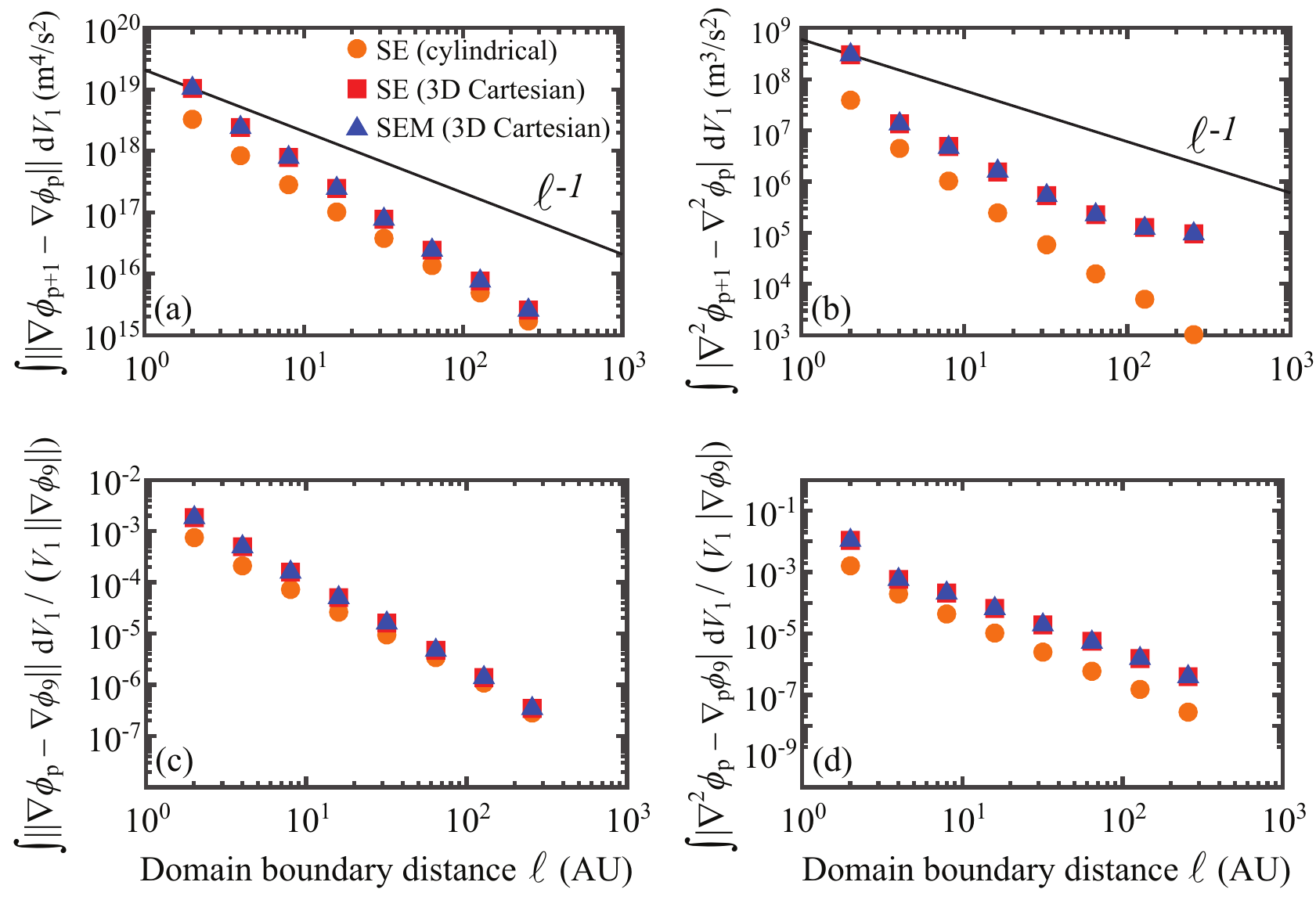}
\caption{
\label{fig:distant_boundary_convergence}
Results of convergence tests for increasing domain boundary distance $\ell \, \textrm{AU}$ for the Galileon force and Laplacian fields corresponding to the Sun-Earth (SE) (cylindrical), Sun-Earth (SE) (3D Cartesian) and Sun-Earth-Moon (SEM) (3D Cartesian) solutions. The largest domain boundary distance is $\ell = 2^{\textrm{p}=9} \, \textrm{AU} = 512 \, \textrm{AU}$. Additional details can be found in Section \ref{sec:numerval:bcs}. Shown for comparison in (a) and (b) is a line with a fall off rate of $1/\ell$. (a) Log-log plots of $\int\|\nabla \dimphi_{\textrm{p}+1} - \nabla \dimphi_{\textrm{p}}\|dV_1$ $m^4/s^2$. (b) Log-log plots of $\int |\nabla^2 \dimphi_{\textrm{p}+1} - \nabla^2 \dimphi_{\textrm{p}}|dV_1$ $m^4/s^2$. (c) Log-log plots of $\int \|\nabla \dimphi_{\textrm{p}} - \nabla \dimphi_{9}\| dV_1$ normalized by $V_1 \|\nabla \dimphi_{9}\|$. (d) Log-log plots of $\int |\nabla^2 \dimphi_{\textrm{p}} - \nabla^2 \dimphi_{9}| dV_1$ normalized by $V_1 |\nabla^2\dimphi_{9}|$.
}
\end{figure*}
%
\begin{figure}[!h]
\centering
\includegraphics[scale=0.9]{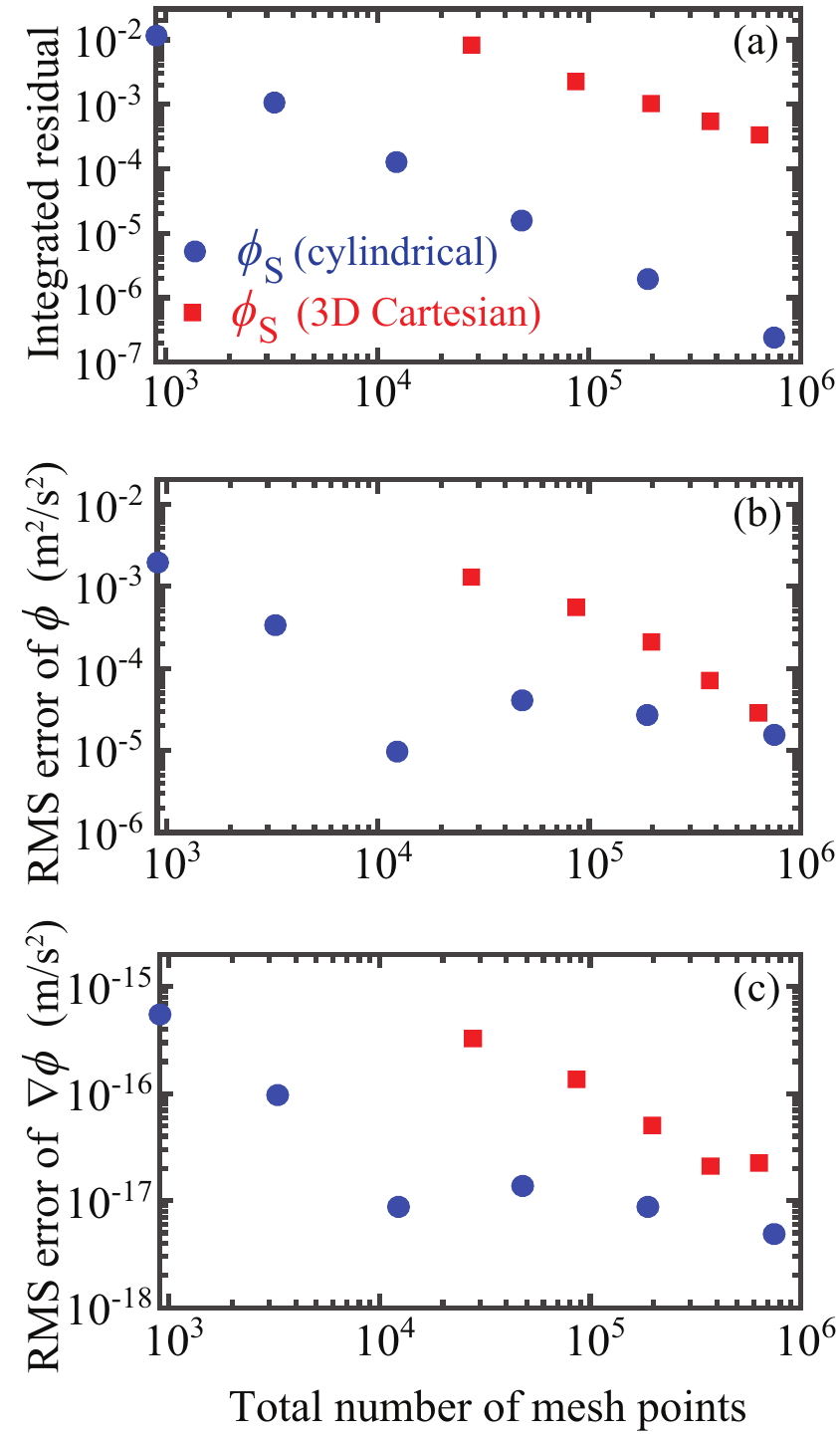}
\caption{
\label{fig:meshconvergence}
Mesh refinement study quantifying the difference between the single-body Sun solutions, $\dimphiS$ and $\nabla \dimphiS$, and the exact analytical result given by Eq. (\ref{eqn:residual}), namely $\dimphi_\textrm{theor}$. Results were carried out with a  cylindrical and cubic domain of boundary distance 64 AU and volume $V_6$. Since the computational volume for each geometry was held constant, increasing number of mesh points reflects smaller mesh lengths. (a) Integrated residual value $d^{-3}\int (\scR^{(n)}[\ndphi])^2 dV_6$. (b) Root mean square error (RMS) of $\dimphi$ given by $[\int (\dimphi_S -\dimphi_{\textrm{theor}})^2 dV_6 / V_6]^{1/2}$. (c) RMS value of $\nabla \dimphi$ given by $[\int \|\nabla \dimphi_S - \nabla \dimphi_{\textrm{theor}}\|^2 dV_6 /V_6]^{1/2}$.
}
\end{figure}

Based on these results, we chose a domain boundary distance of 64 AU, measured from the midpoint of the axis connecting the Sun-Earth bodies, as the standard domain boundary distance for the main computations presented in the body of this work. The improved convergence seen in Fig. \ref{fig:distant_boundary_convergence} for the SE (cylindrical) system is likely due to the finer meshes used there. In particular, when comparing the slope of the curves in Fig. \ref{fig:distant_boundary_convergence}(b) connecting the final two points, we find for the SE (3D Cartesian) system yields a value slightly greater than $-1$ while the SE (cylindrical) yields a value closer to $-1.5$, indicating more rapid convergence.

\subsection{Convergence with mesh refinement}
\label{sec:meshconvergence}

A mesh refinement study was conducted comparing the difference between the numerical solution $\dimphi_\textrm{S} (\vec{r})$ and the analytic solution for the Galileon field $\dimphi_\textrm{theor}(\vec{r})$ of the single-body Sun system given by Eq. (\ref{eqn:spherk0}). Both cylindrical and 3D Cartesian volumes were used with boundary distance $\ell = 64$ (i.e. $V_6 \, \textrm{AU}^3$). The cylindrical volume contained 26 submeshes and the 3D Cartesian volume contained 32 submeshes. For the cylindrical coordinate system, the underlying rectangular mesh elements contained $(n+1)\times(2n+1)$ mesh points per side for $n=$ 4, 8, 16, 32, 64 and 128. For the 3D Cartesian system, the underlying cubic mesh elements contained  $(2n+1)$ mesh points per side for $n=$ 4, 6, 8, 10 and 12. The total number of mesh points was therefore approximately $m \times (2n+1)(n+1)$ for the cylindrical volume and $m \times (2n+1)^3$ for the 3D cubic volume. These numbers are not exact because some points are shared between submeshes and some submeshes contained $(4n+1)$ points per side instead of $(2n+1)$ for the reasons described in Section \ref{sec:numerical_method:meshes}.

The results in Fig. \ref{fig:meshconvergence} for either geometry at constant volume confirm that the integrated residual error decreases monotonically with increasing mesh refinement as shown in (a), indicating that the numerical results approach the analytical results as the total number of mesh points is increased. The root-mean-square (RMS) error for $\ndphi$ in (b) and $\nabla\ndphi$ in (c) also decreases, though not entirely monotonically. In particular, two somewhat odd features are apparent. Firstly, the simulations conducted within a cylindrical volume exhibit a dip of about an order of magnitude at the third mesh refinement step. This is likely a spurious effect, perhaps reflecting that the distribution of points at that mesh size better captures the spherical contours about the  Sun center. Regardless, the error continues to decrease monotonically upon further mesh refinement. Secondly, the RMS error of $\nabla \ndphi$ for the 3D Cartesian system increases slightly at the final mesh refinement step, while that of the cylindrical system continues to drop. This suggests that the simulation results may become more accurate far from the Sun and slightly less accurate near the Sun. However, this behavior may also arise from numerical issues in connection with the fact that the linear problem was solved approximately by using the MATLAB \texttt{bicg} biconjugate gradient solver instead of the direct linear solver. Additional tests conducted using even finer meshes will help resolve this issue.

\subsection{Study of computational times}
\label{comptime}
\begin{figure}[!h]
\centering
\includegraphics[scale=0.9]{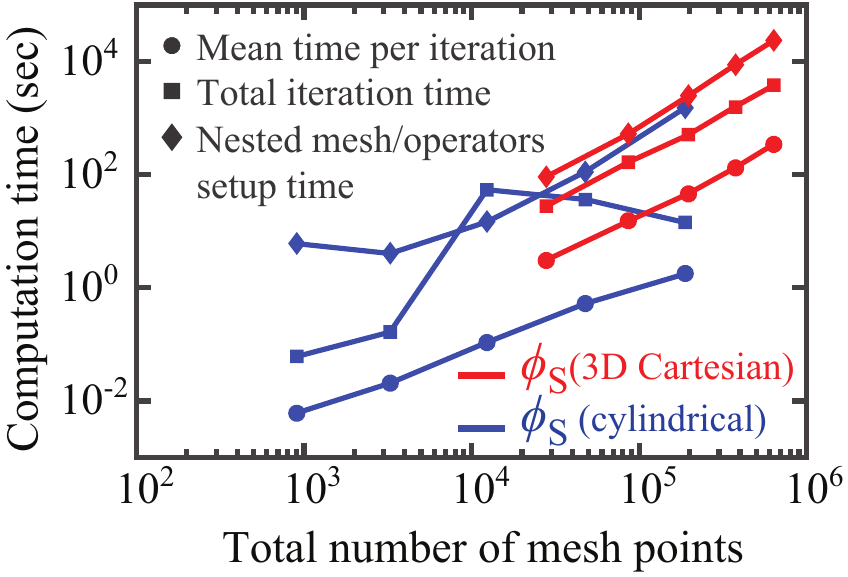}
\caption{
\label{fig:benchmarking}
Mean time per iteration, total iteration time and nested mesh/differential operator setup time for the single-body potential field $\dimphi_\textrm{S}(\vec{r})$ with increasing total number of mesh points. The volume of the cylindrical and cubic domains was $V_6$. Additional details given in Section \ref{comptime}.
}
\end{figure}

Simulations were also conducted to quantify the mean time per iteration, total iteration time and time for constructing nested mesh differential operators for the single-body field $\dimphi_\textrm{S}(\vec{r})$ by increasing the number of total mesh points with a cylindrical and a cubic domain referenced to a volume $V_6$. The computations were performed on a Dell Power Edge R430 server with two 10-core Intel Xeon E5-2630 v4 2.2GHz processors and 112GB of RAM [including 25M Cache, 8.0 GT/s QPI, Turbo, HT, 10C/20T, Max Mem 2133 MHz]. Our software code was not parallelized although some matrix operations in MATLAB automatically run in parallel across multiple cores. The initial trial function for this study was chosen to be the analytic solution given by Eq. (\ref{eqn:spherk0}), which of course is not an exact solution once discretized. The total number of mesh points ranged from 903 for the cylindrical domains with the coarsest meshes to $635,941$ for the cubic domains with the finest meshes.

The results in Fig. \ref{fig:benchmarking} show that for the cylindrical volume, the mean time per iteration scales approximately linearly with the total number of mesh points, while the cubic volume scales somewhere closer to a quadratic. The total iteration time in either case does not increase monotonically due to the variable number of iterations required for the residual to cease to decrease. The time required to set up the initial nested meshes and discrete differential operators appears to scale somewhat between linear and quadratic for both geometries. The results in the main body of this paper were obtained with 47,985 mesh points for the cylindrical volumes with a radius measuring $2^6 = 64 $ AU, resulting in computation times on the order of one or two minutes, and 374,411 mesh points for the cubic volumes of side half-length measuring $2^6 = 64 $ AU, resulting in computation times on the order of thirty minutes.

\acknowledgments
NCW gratefully acknowledges financial support from a 2017 NASA Space Technology Research Fellowship (80NSSC17K0139). This research was carried out in part at the Jet Propulsion Laboratory, California Institute of Technology (Caltech), under a contract with the National Aeronautics and Space Administration (80NM0018D0004). SMT and NCW wish to thank Dr. Peter Thompson for his efforts in administering and upgrading the computing cluster used for the simulations. J\'er\^ome Gleyzes, Jason Rhodes, Olivier Dor\'e, and Eric Huff are also acknowledged for their support of the study and valuable discussions of dark energy and modified gravity.

\bibliography{vainshtein_refs}

\end{document}